\documentclass[sigconf]{acmart}

\AtBeginDocument{%
  \providecommand\BibTeX{{%
    \normalfont B\kern-0.5em{\scshape i\kern-0.25em b}\kern-0.8em\TeX}}}

\copyrightyear{2023}
\acmYear{2023}
\setcopyright{acmlicensed}\acmConference[WWW '23]{Proceedings of the ACM
Web Conference 2023}{May 1--5, 2023}{Austin, TX, USA}
\acmBooktitle{Proceedings of the ACM Web Conference 2023 (WWW '23), May
1--5, 2023, Austin, TX, USA}
\acmPrice{15.00}
\acmDOI{10.1145/3543507.3583209}
\acmISBN{978-1-4503-9416-1/23/04}

\usepackage{pifont}
\usepackage{amsfonts, amsthm}
\usepackage[ruled,vlined,linesnumbered]{algorithm2e}
\usepackage{color, colortbl}
\usepackage{enumitem,multirow,graphicx,subcaption,multicol,lipsum,float,adjustbox}
\newlength{\textfloatsepsave}  

\begin{document}


\title{Distillation from Heterogeneous Models \\for Top-K Recommendation}

\author{SeongKu Kang$^1$$^\ddag$, Wonbin Kweon$^1$, Dongha Lee$^2$, Jianxun Lian$^3$, Xing Xie$^3$, Hwanjo Yu$^1$}
\affiliation{%
   \institution{$^1$Pohang University of Science and Technology (POSTECH), South Korea}
   \institution{$^2$Yonsei University, South Korea\,\,\,\,\,\,$^3$Microsoft Research Asia, Beijing, China}
   \country{}
   \{seongku, kwb4453, hwanjoyu\}@postech.ac.kr, donalee@yonsei.ac.kr, jianxun.lian@outlook.com, xingx@microsoft.com
}
\authornote{Corresponding author\\
$^\ddag$Work is done during the internship at Microsoft.}

\def \authors{SeongKu Kang, Wonbin Kweon, Dongha Lee, Jianxun Lian, Xing Xie, Hwanjo Yu}

\renewcommand{\shortauthors}{SeongKu Kang et al.}

\begin{abstract}
Recent recommender systems have shown remarkable performance by using an ensemble of heterogeneous models.
However, it is exceedingly costly because it requires resources and inference latency proportional to the number of models, which remains the bottleneck for production.
Our work aims to transfer the ensemble knowledge of heterogeneous teachers to a lightweight student model using knowledge distillation (KD), to reduce the huge inference costs while retaining high accuracy.
Through an empirical study, we find that the efficacy of distillation severely drops when transferring knowledge from heterogeneous teachers.
Nevertheless, we show that an important signal to ease the difficulty can be obtained from the teacher's training trajectory.
This paper proposes a new KD framework, named \proposed, that guides the student model by transferring easy-to-hard sequences of knowledge generated from the teachers' trajectories.
To provide guidance according to the student's learning state, \proposed uses \textit{dynamic knowledge construction} to provide progressively difficult ranking knowledge and \textit{adaptive knowledge transfer} to gradually transfer finer-grained ranking information.
Our comprehensive experiments show that \proposed significantly improves the distillation quality and the generalization of the student model.

\end{abstract}


\newcommand{\proposed}{HetComp\xspace}
\maketitle

\section{Introduction}
Recommender system (RS) has been deployed in various applications to facilitate decision-making \cite{he2020lightgcn}.
The core of RS is to provide a personalized ranking list of items to each user.
In the past decades, a variety of models with different architectures and loss~functions, from matrix factorization \cite{BPR} to graph neural networks \cite{he2020lightgcn}, have been studied to generate high-quality ranking lists.
It is known that these heterogeneous models possess different inductive biases that make the model prefer some hypotheses over others \cite{inductive3, MT_KD2}, and accordingly, they better capture certain user/item preferences that better fit the bias of each model \cite{zhu2020ensembled, concf, jointAE}.
As a result, utilizing their multi-faceted knowledge via model ensemble often achieves significantly increased accuracy over a single model \cite{MT_KD2, MT_KD4, zhu2020ensembled, concf, rank_aggregation}.
However, the fundamental limitation is that its computational cost for inference can be many times greater than that of a single model, which makes it impracticable to apply to real-time services.

An increasingly common way to reduce inference latency is to compress a large model into a smaller model via knowledge distillation (KD) \cite{KD}.
KD trains a compact model (student) by transferring the knowledge from a well-trained heavy model (teacher), effectively narrowing the performance gap between them.
Inspired by its huge success in computer vision, KD has been actively studied to compress the ranking model.
Recent \textit{ranking matching} approach \cite{darkrank, DERRD, DCD, reddi2021rankdistil, GCN_distill}
adopts listwise learning that trains the student to emulate the permutations of items (i.e., ranking list) from the teacher.
This approach has shown remarkable performance in many ranking-oriented applications such as top-$K$ recommendation \cite{DERRD, DCD, GCN_distill}, document retrieval \cite{reddi2021rankdistil, CL-DRD}, and person identification \cite{darkrank}.
Nevertheless, they mostly focus on distillation from a homogeneous teacher that has the same model type as the student.
Cross-model distillation from heterogeneous teachers, which have distinct architectures and loss functions, has not been studied well.


\begin{figure*}[t]
\centering
\captionsetup[subfigure]{justification=centering}
\begin{subfigure}[t]{0.22\linewidth}
    \includegraphics[height=4cm]{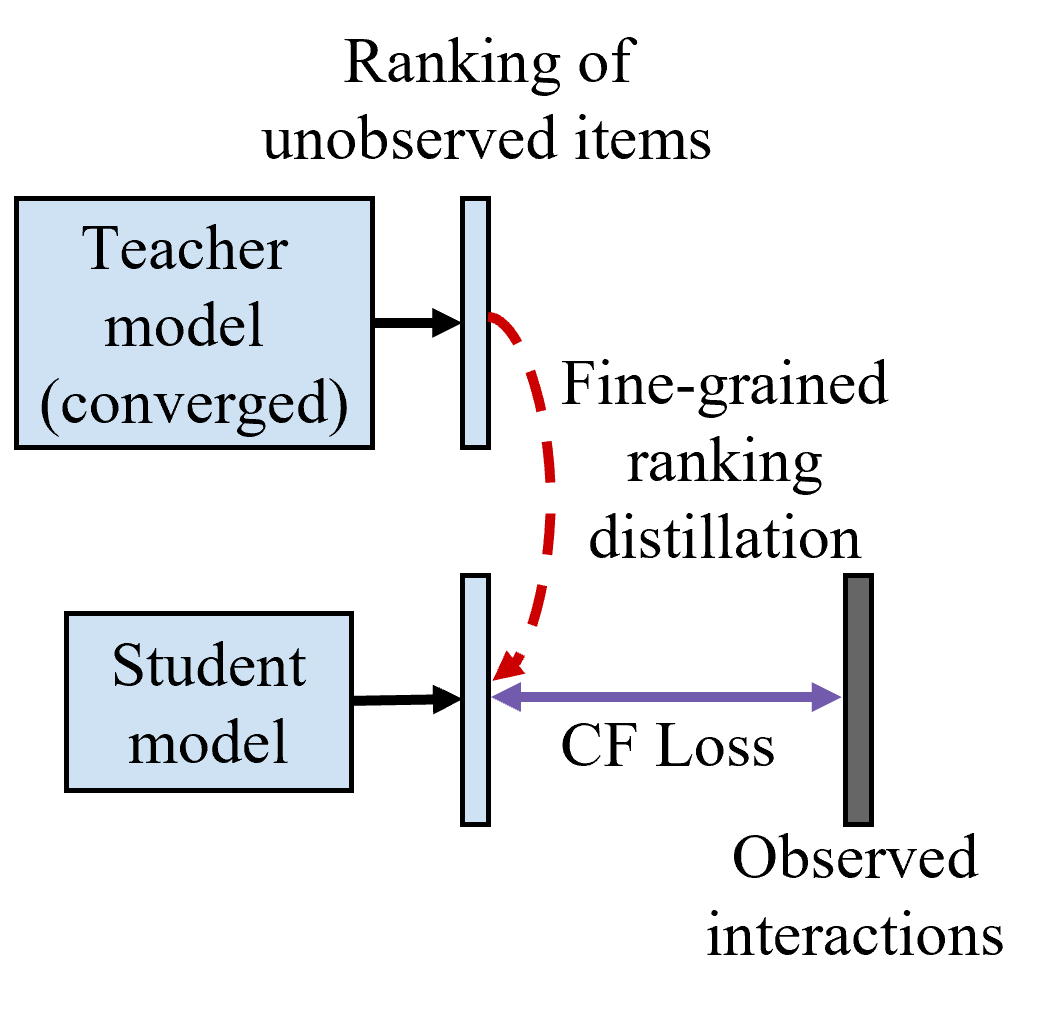}
\end{subfigure}
\hspace{1cm}
\begin{subfigure}[t]{0.65\linewidth}
    \includegraphics[height=4cm]{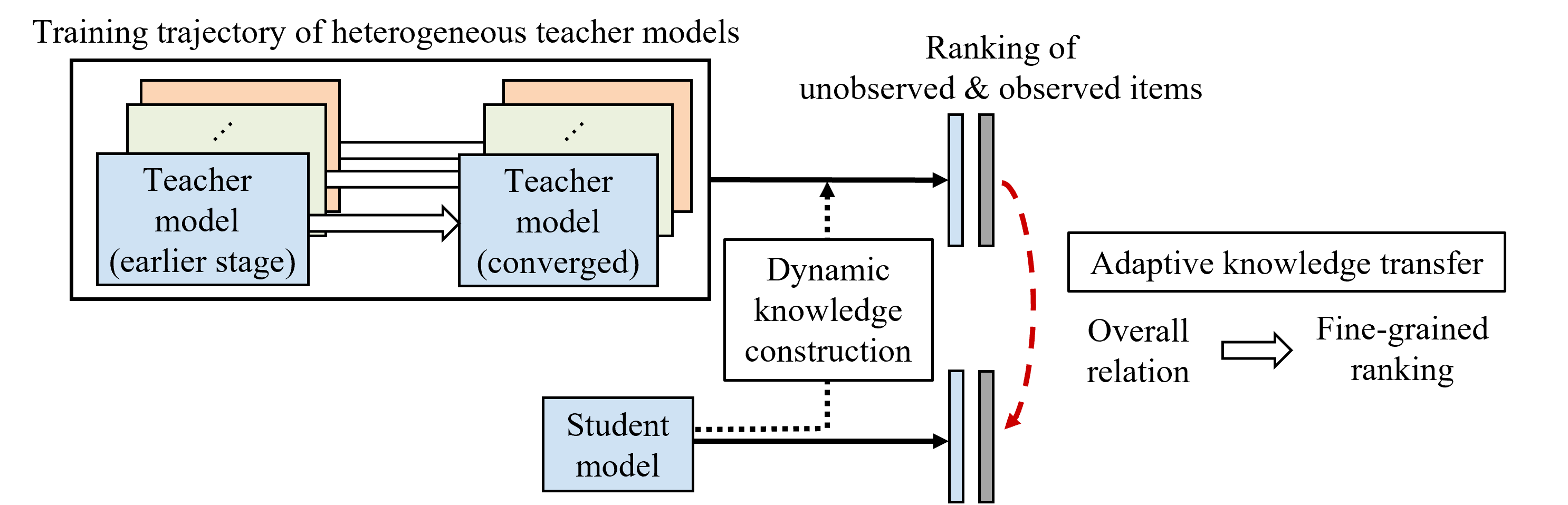}
\end{subfigure}
\vspace{-0.05cm}
\caption{
A comparison of (left) the existing KD approach and (right) our proposed \proposed.
\proposed supervises the student model by using an easy-to-hard sequence of ranking knowledge along with the adaptive distillation objective.
}
\label{fig:intro}
\vspace{-0.4cm}
\Description{A comparison of (left) the existing KD approach and (right) our proposed HetComp. HetComp supervises the student model by using an easy-to-hard sequence of ranking knowledge along with the adaptive distillation objective.}
\end{figure*}

From our analysis, we observe that the efficacy of the prior KD methods significantly drops when transferring knowledge from heterogeneous teacher models, which eventually prevents the student from retaining their ensemble accuracy.
To be specific, we investigate discrepancies between ranking predictions from the student and the teacher, and we observe that the student learning from the heterogeneous teacher has a notably large discrepancy compared to the student learning from the homogeneous teacher.
Fortunately, we find that an important clue that can help to reduce the discrepancy can be obtained from the teacher's intermediate training states;
the students learning from the teacher's earlier training states have considerably lower discrepancies than the student learning from the final converged teacher, regardless of the type of teacher model.
This indicates that during the teacher's training, the teacher's knowledge gets gradually harder for the student to learn.
Indeed, we empirically show that the teacher's predictions become increasingly complex, as the latter predictions contain more diversified and individualized item~rankings.

Motivated by the observation, our key idea for improving the distillation efficacy is to dynamically transfer knowledge in an easy-to-hard sequence by using the teacher models' training trajectories (Figure \ref{fig:intro}).
As the teachers' knowledge is gradually becoming harder during their training, we aim to supervise the student model to gradually follow such sequences for easing the difficulties of learning from the converged teachers.
Our approach is based on the easy-to-hard learning paradigm \cite{jiang2015self, curriculum, Curri_survey} which has been extensively studied in various fields of machine learning.
It trains the model by using easy samples first and progressively more difficult samples so that the model can leverage previously seen concepts to ease the acquisition of more difficult ones \cite{jiang2015self, curriculum}.
Following the idea, we start distillation with relatively easy knowledge from the teachers' early training stage, then gradually increase the difficulty along the teachers' trajectories,  reducing the huge discrepancy incurred when learning from heterogeneous teachers.


In this work, we present a new KD framework, termed as \proposed, that effectively compresses the valuable but difficult ensemble knowledge of heterogeneous models, generating a lightweight model with high recommendation performance.
\proposed concretizes the easy-to-hard distillation in the following aspects:
\begin{itemize}[leftmargin=*] \vspace{-\topsep}
    \item \textbf{What to transfer}: \proposed supervises the student model via \textit{dynamic knowledge construction} which provides the easy-to-hard sequence of permutations considering the student's learning state.
    It first identifies proper knowledge from the trajectory of each teacher, and then dynamically constructs the target knowledge to guide the student model.
    
    \item \textbf{How to transfer}: \proposed uses \textit{adaptive knowledge transfer} which adjusts the distillation objective according to the student's learning state.
    It trains the student model to first focus on the overall relations in the target permutations, and gradually move on to learning to rank the fine-grained orders of the preferable items.
    Furthermore, we introduce a new transfer strategy to exploit the knowledge of both observed and unobserved user-item interactions, which has been not considered in prior~works.
\end{itemize}
\vspace{-\topsep}

\noindent
Our contributions are summarized as follows:
\begin{itemize}[leftmargin=*] \vspace{-\topsep}
    \item We reveal the difficulty of the ranking knowledge transfer from heterogeneous models and tackle the issue from the perspective of easy-to-hard distillation, which is new for recommendation.
    
    \item We propose \proposed, a new KD framework that effectively compresses the ensemble of heterogeneous models into a compact model. 
    \proposed can significantly ease the huge computational burdens of the model ensemble while retaining its high accuracy.

    \item We validate the superiority of \proposed by extensive experiments on real-world datasets.
    We also provide a comprehensive analysis of our proposed approach.
\end{itemize}
\vspace{-\topsep}

\section{Related Work}
\label{sec:relatedwork}

\noindent
\textbf{Knowledge Distillation.}
Knowledge distillation (KD) has been actively studied for model compression in various fields \cite{curri_dialog, darkrank, KD, FitNet, xia2022device, TD}.
KD transfers the knowledge captured by a teacher model through large capacity into a lightweight student model, significantly lowering the inference cost while maintaining comparable performance.
Pointing out that the knowledge from a single teacher model is insufficient to provide accurate supervision, many recent studies \cite{MT_KD1, MT_KD2, MT_KD3, MT_KD4, MT_KD5, MT_KD6, zhu2020ensembled} employ multiple teacher models and show great effectiveness in further improving a student model.
Notably, the state-of-the-art methods \cite{MT_KD2, MT_KD4, MT_KD6} exploit heterogeneous teacher models varying in configurations, architectures, loss functions, and many other factors to incorporate their complementary knowledge, which can provide more comprehensive guidance than a single view from a single or homogeneous teacher model.

\noindent
\textbf{Knowledge Distillation for Ranking.}
KD has been also studied for ranking problems.
Many studies \cite{RD, PHR, BD, zhu2020ensembled, chen2022learning} transfer~point-wise importance on each user-item pair (or query-document pair).
However, the point-wise approach cannot consider the relations of multiple items simultaneously, which leads to the limited ranking performance \cite{darkrank, GCN_distill, DERRD}.
Recent methods \cite{DERRD, DCD, GCN_distill, CL-DRD, reddi2021rankdistil, darkrank} formulate the distillation process as a \textit{ranking matching} task.
They utilize the ranking orders from the teacher as supervision and train the student to preserve the teacher's permutation.
By directly transferring the ranking knowledge, this approach has shown state-of-the-art performance in various ranking-oriented applications such as top-$K$ recommendation \cite{DERRD, DCD, IRRRD, GCN_distill}, document retrieval \cite{reddi2021rankdistil, CL-DRD}, and person identification \cite{darkrank}.
Further, the ranking matching approach can be flexibly applied to knowledge transfer between heterogeneous models having distinct output score distributions to which the point-wise approach cannot be~directly~applied~\cite{concf}.

\noindent
\textbf{Easy-to-hard Learning.}
Inspired by the learning process of humans, easy-to-hard learning has been extensively studied in various fields of machine learning \cite{jiang2015self, curriculum, kumar2010self, liu2017easy, macavaney2020training, curri_RS, Curri_survey, wu2020curricula}.
It has been widely used when direct optimization of a non-convex objective function may converge to poor local minima and has been proven to play an important role in achieving a better generalization \cite{curriculum}.
Curriculum learning \cite{curriculum, Curri_survey} trains a model by gradually including data samples in ascending order of difficulty defined by prior knowledge.
On the other hand, self-paced learning \cite{kumar2010self} makes the curriculum dynamically adjusted during the training, usually based on training loss \cite{kumar2010self} or performance on the validation set \cite{curri_RS, xiang2020learning}.
The easy-to-hard learning has been applied to KD to improve the distillation efficacy in computer vision \cite{shi2021follow, RCO} and natural language processing \cite{curri_dialog, CL-DRD}.
\cite{RCO, shi2021follow, cazenavette2022dataset} exploit the teacher's optimization route~to~form a curriculum for the student, \cite{CL-DRD} gradually includes an increasing number of fine-grained document pairs during the~training.

\noindent \textbf{Remarks.}
The existing KD methods for RS focus on distillation from a homogeneous teacher that has the same model type to the student model.
Distillation from heterogeneous teachers, which have distinct architectures and learning objectives to the student model, has not been studied well.
In this work, we show the necessity and difficulty of distilling the ensemble of heterogeneous teachers and apply the easy-to-hard learning to cope with the problem.
Further, the prior KD works with the easy-to-hard learning focus on classification \cite{TA_KD, shi2021follow, RCO} or rely on domain-specific features \cite{curri_dialog}, which makes them hard to apply to the ranking problem and recommender system.
Our work provides a solution tailored to compress ranking models by distilling an easy-to-hard sequence of ranking knowledge considering the student's learning state.


\section{Preliminaries}
\label{sec:preliminary}
\subsection{Problem Formulation}
Let $\mathcal{U}$ and $\mathcal{I}$ denote the user and item sets, respectively.
Given implicit user-item interaction (e.g., click) history, a recommendation model $f: \mathcal{U} \times \mathcal{I} \rightarrow \mathbb{R}$ learns the ranking score of each user-item pair. 
Based on the predicted scores, the recommender system provides a ranked list of top-$K$ unobserved items for each user, called as top-$K$ recommendation.
Given a set of cumbersome teacher models $\mathcal{F} = \{f^1, f^2, ... , f^M\}$, our goal is to effectively compress an ensemble of the teachers into a lightweight student model $f$.
The student model has a significantly reduced computational cost for inference, and thus it is more suitable for real-time services and resource-constrained environments.
We pursue a model-agnostic solution, which enables any kind of recommendation model can be flexibly used for both teacher and student, allowing service providers to use any preferred model according to their environments.

We exploit heterogeneous teacher models with various architectures and loss functions.
In this work, we choose six representative types of models extensively studied for RS: MF (Matrix Factorization) \cite{BPR}, ML (Metric Learning) \cite{CML}, DNN (Deep Neural Network) \cite{NeuMF}, GNN (Graph Neural Network) \cite{he2020lightgcn}, AE (AutoEncoder) \cite{VAE}, I-AE (Item-based AE) \cite{autorec}.
A detailed analysis of the teacher models and their ensemble is provided in~Appendix~\ref{sec:app_MTS}.

\noindent
\textbf{Notations.}
Given a ranked list (i.e., permutation of items) $\pi$, $\pi_k$ denotes the $k$-th item in $\pi$, and $r(\pi, i)$ denotes the ranking of item $i$ in $\pi$ where a lower value indicates a higher position, i.e., $r(\pi, i)=0$ is the highest ranking.
Note that $\pi$ is defined for each user $u$.
For notational simplicity, we omit $u$ from $\pi$ throughout the paper.

\subsection{Ranking Matching Distillation}
\label{sec:rankingKD}
Ranking matching distillation \cite{DERRD, DCD, GCN_distill, darkrank, reddi2021rankdistil} trains the student model to emulate the teacher's permutation.
A dominant strategy is to associate a probability with every permutation based on the Plackett-Luce model \cite{marden1996analyzing}, then train the student to maximize the likelihood of the teacher's permutations \cite{xia2008list-wise}.
Given the teacher's permutation $\pi^t$ on a user $u$, the recent studies \cite{reddi2021rankdistil, DCD, DERRD} match the ranking of top-ranked items ($P$) while \textit{ignoring} the ranking of the remaining items ($N$).
The listwise KD loss is defined as the negative log-likelihood of permutation probability of $[P;N]$~as~follows:
\begin{equation}
\mathcal{L}_{F}(P, N) = - \log \prod_{{k}={1}}^{|P|} 
 \frac{\exp \, f(u, P_k)}
  {\sum_{{j}={k}}^{|P|}  {\exp} \,  f(u, P_j) + {\sum_{{l}={1}}^{|N|} {\exp} \, f(u, N_l)}}
\label{Eq:soft_listmle}
\end{equation}
$P$ and $N$ are mostly chosen to be a few top-ranked items and a subset of items randomly drawn from the numerous remaining items, respectively \cite{DERRD, reddi2021rankdistil}.
By minimizing the loss, the student model learns to preserve the fine-grained orders in $P$, while penalizing items in $N$ below the lowest ranking of items in $P$.
It is worth noting that the orders of items within $N$ are not necessarily preserved.

\subsection{Study on Ranking Knowledge Distillation}
We present our analysis showing the difficulty of our task that compresses the ensemble knowledge of heterogeneous teacher models.
Further, we show that a clue that helps to ease the difficulty can be obtained from the teachers' intermediate training states.
Here, we use MF with embedding size 6 as the student model, and all teacher models have embedding size 64. 
Similar results are also observed with other types of students.
We train the student solely with distillation (Eq.\ref{Eq:soft_listmle}).
Please refer to Sec.\ref{sec:experimentsetup} for the detailed setup.

\vspace{-0.05cm}
\subsubsection{\textbf{Discrepancy.}}
Since recommendation accuracy only reveals the efficacy of KD in an indirect way, we introduce a new metric for directly assessing how closely the student model learns the teacher's permutation $\pi^t$.
Let $\pi$ denote the permutation predicted by the student model.
We define the discrepancy between~$\pi$~and~$\pi^t$~by
\begin{equation}
\begin{aligned}
    D@K(\pi, \pi^{t}) = 1 - NDCG@K(\pi, \pi^{t}),
\end{aligned}
\end{equation}
where $D@K(\pi, \pi^{t})=0$ indicates the student model perfectly preserves top-$K$ ranking of $\pi^t$.
$NDCG$ is a widely used listwise ranking evaluation metric \cite{CofiRank}.
Here, we consider $\pi^t$ as the optimal ranking. 
\begin{equation}
\begin{aligned}
\hspace{-0.1cm}
\textit{\small{NDCG@K}}(\pi, \pi^t)=\frac{\textit{\small{DCG@K}}(\pi)}{\textit{\small{DCG@K}}\left(\pi^t\right) },\,  \textit{\small{DCG@K}}(\pi)=\sum_{k=1}^{K} \frac{2^{y_{\pi_k}}-1}{\log (k+1)}
\end{aligned}
\end{equation}
The relevance of each item $i$ (i.e., $y_i$) is mostly defined as ratings (for explicit feedback) or binary values (for implicit feedback).
To put a higher emphasis on a top-ranked item in $\pi^t$, we use the parametric geometric distribution \cite{rendle2014improving}, i.e., $y_i = \exp(-r(\pi^{t}, i) / \lambda)$ if $i$ is within the top-$K$ of $\pi^t$, otherwise 0.
$\lambda \in \mathbb{R}^+$ is the hyperparameter that controls the sharpness of the distribution.

\vspace{-0.05cm}
\subsubsection{\textbf{Observations and analyses.}}
We train the student model (MF) by distillation from various ranking supervisions and analyze the discrepancy between the student and the supervision.
In Table~\ref{tab:pre}, (a) denotes a homogeneous teacher, which has the same model type (i.e., MF) to the student, as used in most previous work.
(b) and (c) denote the ensemble of six homogeneous teachers with different initialization and the ensemble of six heterogeneous teachers, respectively.
`NLL' denotes the negative log-likelihood of the student model for the given supervision (Eq.\ref{Eq:soft_listmle}) where a lower value implies the student better emulates the given supervision.
We compute the metrics for each user and report the average value.
In Table \ref{tab:pre}, we observe that \textit{compressing (c) incurs a notably large discrepancy compared to compressing (a) and (b).}
In other words, the efficacy of distillation is severely degraded when we transfer the ensemble knowledge of heterogeneous teachers.
This is an interesting observation showing that items' ranking orders in permutations bring a huge difference to learning~difficulty.

\begin{table}[t]
\caption{Discrepancy to the given supervision after KD.
}
\label{tab:pre}
\small
\centering
\renewcommand{\arraystretch}{0.6}
\renewcommand{\tabcolsep}{0.8mm}
\resizebox{1.01\linewidth}{!}{
\begin{tabular}{c|lc|ccc}
\hline
\multirow{2}{*}{\textbf{Dataset}} & \multicolumn{2}{c|}{\textbf{Supervision (Teacher)}} & \multicolumn{3}{c}{\textbf{Discrepancy}} \\
\cline{2-6}
 & \multicolumn{1}{c}{\textbf{Type}} & \textbf{Recall@50} & \textbf{D@10} & \textbf{D@50} & \textbf{NLL} \\
\hline\hline
 & (a) Single-teacher (MF) & 0.2202 & 0.6640 & 0.5167 & 0.5805 \\
Amusic & (b) Ensemble (MF)  & 0.2396 & 0.6699 & 0.5162 & 0.6048 \\
 & (c) Ensemble (Het)  & 0.2719 & 0.7417 & 0.5958 & 0.7206\\ \hline
 & (a) Single-teacher (MF) & 0.2604 & 0.5101 & 0.3716 & 0.5962 \\ 
CiteULike  & (b) Ensemble (MF) & 0.2763 & 0.5373 & 0.3910 & 0.5977 \\
 & (c) Ensemble (Het) & 0.3144 & 0.6983 & 0.5269 & 0.6906 \\ 
\hline
\end{tabular}}
\vspace{-0.4cm}
\end{table}

\begin{figure}[t]
\centering
\hspace{-0.15cm}
    \includegraphics[width=0.51\linewidth]{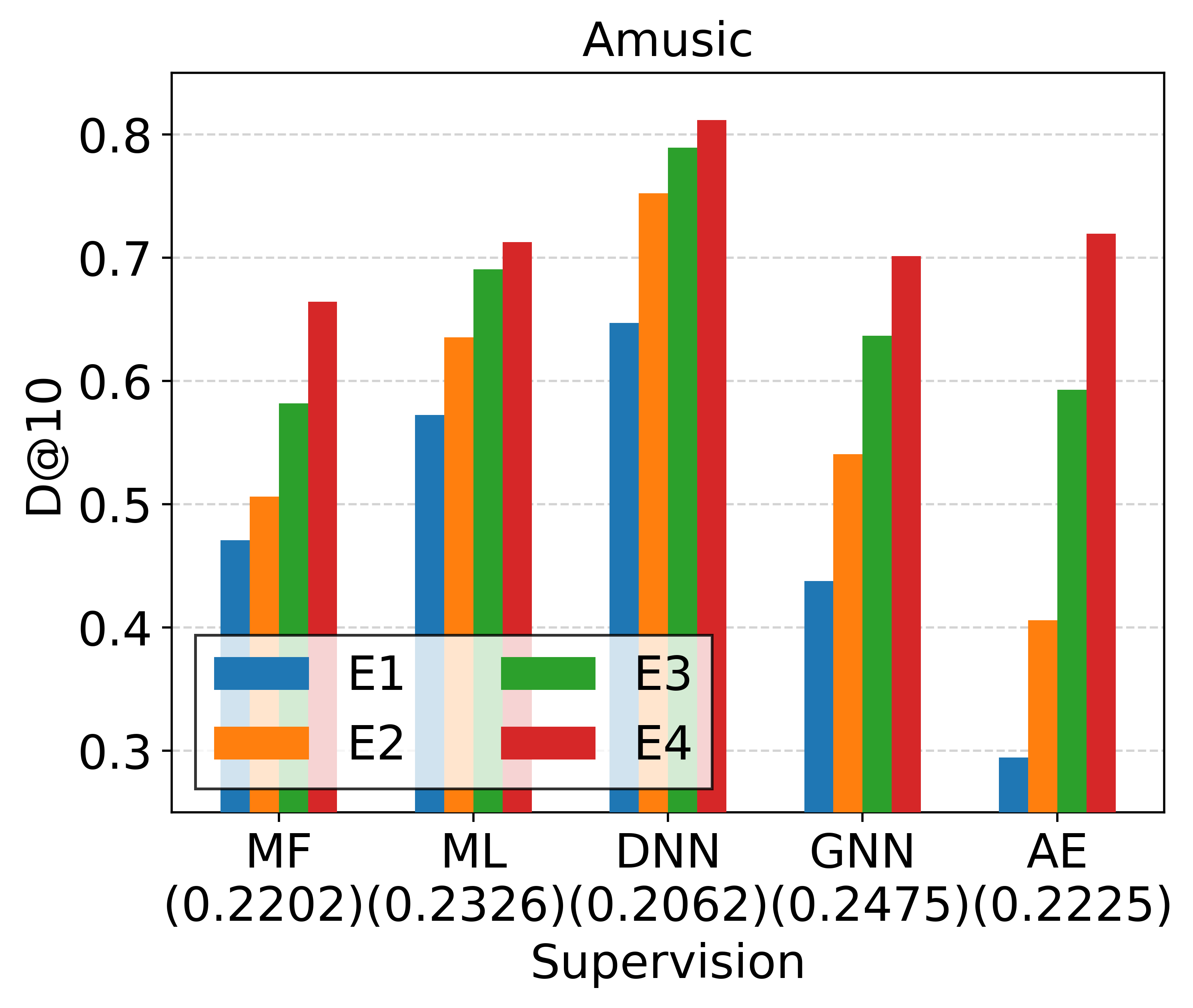}
    \hspace{-0.2cm}
    \includegraphics[width=0.51\linewidth]{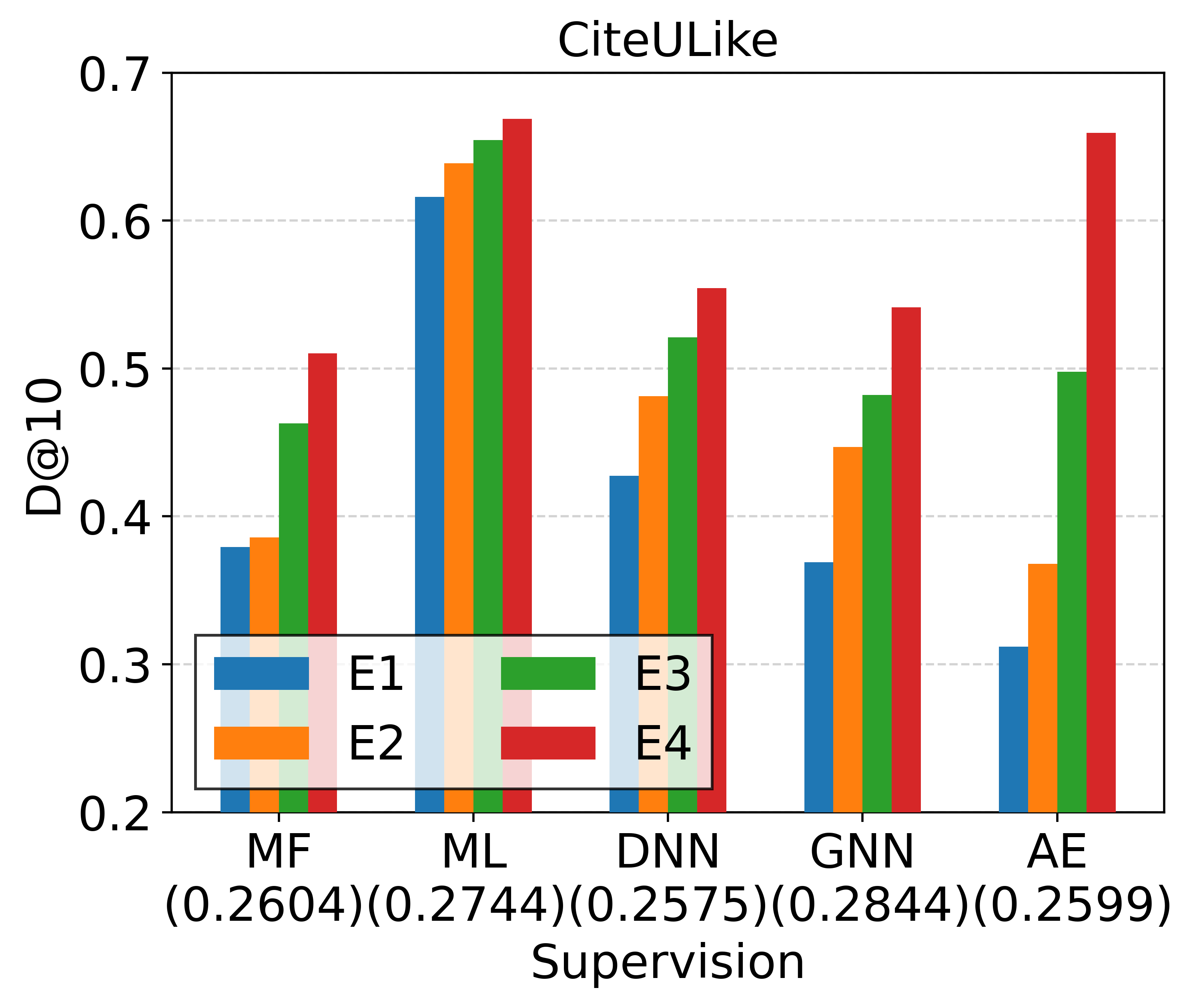}\hspace{-0.15cm}
    \caption{Discrepancy to various supervisions from intermediate training states (E1/E2/E3) and the final converged state (E4) of each teacher model. 
    We also annotate Recall@50 of each converged teacher model below the x-axis.
    }
    \label{fig:pre}
\Description{Discrepancy to various supervisions from intermediate training states (E1/E2/E3) and the final converged state (E4) of each teacher model. We also annotate Recall@50 of each converged teacher model below the x-axis.}
\end{figure}

To investigate where such a large discrepancy originated from, we analyze the cases of learning from each teacher in (c).
We independently train the student model (MF) with distillation from each converged teacher and its intermediate training states\footnote{For each teacher, we use 4 training states (or checkpoints), from E1 to E4, each of which corresponds to the checkpoint at 25\%, 50\%, 75\%, and 100\% of the converged~epochs.
I-AE shows similar results to AE, and its results are omitted due to the limited space.
}.
Then, we present the discrepancy to each supervision in Figure \ref{fig:pre}.
We observe that \textit{compared to the case of the homogeneous model (i.e., MF with E4), distillation from a heterogeneous model (i.e., others with E4) consistently incurs a larger discrepancy.}
It is known that heterogeneous models possess different inductive biases that make the model prefer some hypotheses over others \cite{inductive3, MT_KD2, concf}.
In this sense, learning from heterogeneous teachers can be particularly challenging since it needs to learn relationships that do not fit well with the student model. 
Interestingly, higher teacher accuracy does not necessarily result in a larger discrepancy.
For example, on the CiteULike dataset, GNN achieves higher accuracy and AE achieves comparable accuracy compared to MF.
However, the discrepancy is much higher in the case of learning from AE.

On the one hand, \textit{the discrepancy gradually gets larger during the teachers' training (from E1 to E4), regardless of the model type.}
That is, the teachers' knowledge gets more difficult to emulate during their training process.
As will be shown in Sec.\ref{subsec:studyH}, a model tends to learn overall patterns first, then gradually learns personalized~preferences.
We also show that teachers' knowledge becomes increasingly complex during the training, as the latter predictions contain more diverse and individualized item rankings for each user.
These observations motivate us to utilize the teachers’ training trajectories as a natural curriculum for the student model.


\section{METHODOLOGY}
\label{sec:method}
\proposed (\textbf{Het}erogeneous model \textbf{Comp}ression for RS) framework supervises the student model using the teachers' training trajectories, based on the idea of easy-to-hard learning.
\proposed consists of the two major components designed for easy-to-hard distillation:
\begin{itemize}[leftmargin=*] \vspace{-\topsep}
    \item \textbf{(Sec.\ref{subsec:dkc}) Dynamic knowledge construction} dynamically generates knowledge with appropriate difficulty, considering the student's learning state.
    It first identifies \textit{proper} knowledge from the trajectory of each teacher and constructs target knowledge to guide the student model.
    This process is applied in a personalized manner to transfer each user's recommendation result considering their different learning difficulties.
    
    \item \textbf{(Sec.\ref{subsec:KT}) Adaptive knowledge tranfer} adjusts the distillation objective according to the student's learning state.
    It trains the student to first focus on the overall relations in the target permutations and gradually move on to learning to rank the fine-grained orders of the preferable items.
    Furthermore, we propose a new strategy to transfer the knowledge of both observed and unobserved interactions which is ignored in the~previous~works.
\end{itemize}
\vspace{-\topsep}
The overall training process of \proposed is provided in Sec.\ref{subsec:tp}.

\subsection{Dynamic Knowledge Construction}
\label{subsec:dkc}
The goal of our knowledge construction is to generate target knowledge to be transferred from the teachers' training trajectories,~based on the student's state.
This process has to meet the following~desiderata:
(1) it should reflect the different learning difficulties of knowledge from each teacher model and each user;
our analysis shows that the learning difficulty varies across teacher models, 
and further, ranking with respect to each user naturally has different difficulties.
(2) it should work in a model-agnostic manner so that it can handle any kind of teacher/student model.
For this reason, we avoid using statistics that differ for each model type (e.g., training loss) for deciding on knowledge to be transferred.
We also avoid using a performance metric on additional held-out data, as it is difficult to obtain enough interactions of each user due to the~high~sparsity.

Let $\mathcal{T}=\{T^x\}_{x \in \{1,...,M\}}$ denote the teachers' training trajectories.
For each teacher $f^x$, we use its predictions (i.e., rankings of unobserved items) at $E$ different training states\footnote{In this work, we set $E$ as 4 and evenly distribute them across each teacher's trajectory.
We empirically found that $E$ hardly affects the final performance as long as they are well distributed over the teacher training process (Appendix \ref{app:cost}).}, i.e., $T^x = [\pi^{x, 1}, ..., \pi^{x, E}]$.
The last permutation $\pi^{x, E}$ corresponds to the final prediction after convergence.
Our analysis shows that the teacher's knowledge is getting harder during its training, so that $\pi^{x,e}$ is easier to emulate than $\pi^{x,e+1}$.
So, we start from $\pi^{x,1}$ and then gradually move to $\pi^{x, E}$.
We use $v$ to denote the $M$-dimensional selection vector where each element $v[x] \in \{1,...,E\}$ indicates which training state of the teacher $f^x$ is currently selected to guide the student model.
During the training, we identify proper knowledge from each teacher's trajectory $T^x$, then construct the dynamic supervision $\pi^\text{d}$ by consolidating~them~based on $v$.
The overview of the knowledge construction is provided in Algorithm~\ref{algo:dkc}.

\begin{algorithm}[t]
\SetKwInOut{Input}{Input}
\SetKwInOut{Output}{Output}
\Input{Teachers' trajectories $\mathcal{T}$, student model $f$, current selection variable $v$ with discrepancies $d$}
\Output{New dynamic target permutation $\pi^\text{d}$}
\BlankLine
Predict $\pi$ by student model $f$  

\ForEach{teacher $\,x \in \{1,...,M\}$}{
    \If{$(v[x] < E)$ \, $\operatorname{\mathbf{and}}$ \, $(\gamma^x > \alpha)$}{
        $v[x] = v[x] + 1$\\ 
        $d[x] = D@K(\pi, \pi^{x,\, \min(v[x]+1,\, E)})$
    }
}
Generate $\pi^\text{d} = g(\{\pi^{x, v[x]}\}_{x \in \{1,...,M\}})$
\caption{Dynamic Knowledge Construction}
\label{algo:dkc}
\end{algorithm}


To control the difficulty of distillation, we use the discrepancy between ranking predictions from the student and teachers.
Our key idea is to use the earlier predictions to ease the learning difficulty of the latter predictions.
For each teacher trajectory $T^x$, we keep tracking the discrepancy to the permutation from the next training state $D@K(\pi, \pi^{x, (v[x]+1)})$, and move to the next state if the discrepancy decreases to a certain degree.
To this end, we define the relative discrepancy ratio as follows:
\begin{equation}
\begin{aligned}
    \gamma^x = \frac{d[x]}{D@K(\pi, \pi^{x, (v[x]+1)})},
\end{aligned}
\end{equation}
where $d[x]$ denotes the \textit{initial} discrepancy to $\pi^{x, (v[x]+1)}$ computed when the student \textit{begins to} learn $\pi^{x, v[x]}$.
Note that $d[x]$ is treated as a constant and $D@K(\pi, \pi^{x, (v[x]+1)})$ decreases as the student evolves during the training.
$\gamma^x$ reveals the degree to which the learning difficulty of $\pi^{x, (v[x]+1)}$ is lowered.
Then, we employ a greedy strategy that moves to the next teacher state if $\gamma^x$ becomes larger than the threshold $\alpha$.
$\alpha \geq 1$ is a hyperparameter controlling the transition speed where a lower value incurs a faster transition.
Our greedy strategy based on the discrepancy provides an efficient curriculum considering both the student's learning state and the varying difficulties of the teachers.
Also, as the discrepancy can be measured for any model, it works in~a~model-agnostic~manner.

After updating the selection variable $v$, we generate the target permutation by 
\begin{equation}
\pi^\text{d} = g(\{\pi^{x, v[x]}\}_{x \in \{1,...,M\}})
\label{eq:pid_ensemble}
\end{equation}
where $g$ is the ranking ensemble function to consolidate the permutations.
Here, various ensemble techniques \cite{rank_aggregation}, from a simple averaging to a more sophisticated one with learnable importance, can be flexibly used in \proposed.
In this work, we use a simple technique, which leverages the consistency of prediction as importance, widely used in recent work \cite{NS_std, concf} (Appendix \ref{app:ensemble_technique}).

Once $v[x]$ equals $E$ for all $x$, the knowledge becomes equivalent to the final ensemble knowledge used in the conventional KD.
However, the key difference is that the student model is now more capable of learning from the more difficult knowledge.

\subsection{Adaptive Knowledge Transfer}
\label{subsec:KT}
We present how we transfer the dynamically-constructed permutation $\pi^\text{d}$ to the student.
We first introduce our distillation objective which is adaptively adjusted according to the student's learning state in Sec \ref{subsubsec:ado}.
Then, we explain our strategy to transfer ranking knowledge of both observed and unobserved items in Sec \ref{subsubsec:obs}.

\subsubsection{\textbf{Adaptive distillation objective}}
\label{subsubsec:ado}
Considering the target permutation varies during the student's training, learning the detailed ranking in the earlier phase is not only unnecessary but also daunting.
We first train the student model to learn the overall relations in the target permutation by $\mathcal{L}_{O}$.
Then, once the target permutation is constructed from the final converged predictions (i.e., $v[x]$ equals $E$, $\forall x$), we move on to learning to rank the fine-grained ranking orders by $\mathcal{L}_{F}$.
By modifying Eq.\ref{Eq:soft_listmle}, we define the distillation objective to transfer the overall relations~as~follows:
\begin{equation}
    \mathcal{L}_{O}(P, N) = - \log \prod_{{k}={1}}^{|P|} 
    \frac{\exp \, f(u, P_k)}
    {{\exp} \,  f(u, P_k) + {\sum_{{l}={1}}^{|N|} {\exp} \, f(u, N_l) }},
\label{Eq:overall_listmle}
\end{equation}
It simply pushes the items in $P$ to be ranked higher than items in $N$ without imposing any constraint among the items in $P$.

\subsubsection{\textbf{Transferring knowledge of observed/unobserved items}}
\label{subsubsec:obs}
The prior KD methods \cite{RD, DERRD, DCD} mostly transfer the ranking of unobserved items without considering the ranking of observed items.
We argue that the relative priorities among the observed items are also valuable knowledge of user preference.
A straightforward way to exploit the knowledge of observed items is to transfer a permutation of the whole item set constructed from the teachers.
However, because RS models are mostly trained by collaborative filtering (CF) losses (e.g., BCE \cite{NeuMF}, BPR \cite{BPR} loss) penalizing unobserved items to have lower scores than the observed items, ``relevant but not-yet-observed'' items are likely to have lower ranks compared to their true relevance in the permutation of whole item set.
We observe that if we directly distill the whole item permutation, this causes some not-yet-observed items to be overly penalized in the student's prediction, which hinders the learning of preferences.

We propose a new strategy to exploit both observed and unobserved ranking knowledge effectively.
We \textit{independently transfer} the ranking of observed items and top-ranked unobserved items to prevent such over-penalizing.
Let $P^-$ denote the ranking of top-ranked unobserved items and $N$ denote the remaining unobserved items obtained from $\pi^\text{d}$.
We additionally define $P^+$, the ranking of observed items\footnote{$P^+$ is obtained by the ensemble of the converged teachers on the observed items.
It is worth noting that $P^+$ needs to be generated only once before the student's training as the set of observed items is fixed and $\mathcal{L}_O$ doesn't transfer their detailed ranking.}.
The distillation loss is defined as~follows:
\begin{equation}
    \mathcal{L} = \mathcal{L}_{KD}(P^+, N) + \mathcal{L}_{KD}(P^-, N),
\label{Eq:final_loss}
\end{equation}
where $\mathcal{L}_{KD}$ is either $\mathcal{L}_{O}$ (Eq.\ref{Eq:overall_listmle}) or $\mathcal{L}_{F}$ (Eq.\ref{Eq:soft_listmle}) depending on the student's learning state.
Note that our strategy does not enforce the top-ranked unobserved items ($P^-$) to be located below the observed items ($P^+$), preventing pushing the not-yet-observed items away from the top ranking.
Instead, it enables some unobserved items with high relevance to be naturally mixed with the observed items near the top of the ranking list.
$\mathcal{L}_{KD}(P^+, N)$ is distinguishable from the CF losses in that (1) it transfers the ranking of observed items, and (2) it does not penalize top-ranked unobserved items.

\begin{algorithm}[t]
\small
\SetKwInOut{Input}{Input}
\SetKwInOut{Output}{Output}
\Input{Teachers' trajectories $\mathcal{T}$, student $f$, an update period $p$}
\Output{Trained student model $f$}
\BlankLine
Randomly initialize student model $f$ \\
Initialize selection variables $v_u[x] = 1$ and $d_u[x]  \,\,\,\, \forall x, \forall u$\\
Obtain $P^+_u$ by the ensemble of converged teachers \,\,\,\, $\forall u$\\
\For{$i=1,... ,\text{ }epoch_{max}$  }{
    \ForEach{user $u \in \mathcal{U}$}{
        \If{$(i \text{ } \% \text{ } p == 0)$ \,$\operatorname{\mathbf{and}}$\,
        $\operatorname{\mathbf{not}}$ $(v_u[x] == E\,, \forall x)$}{
            Update $v_u, d_u$, and $\pi^\text{d}_u$ via dynamic knowledge construction (Algorithm \ref{algo:dkc})
        }
        \If{$\operatorname{\mathbf{not}}$ $(v_u[x] == E\,, \forall x)$}{
            $\mathcal{L} = \mathcal{L}_{O}(P^+_u, N_u)$ + $\mathcal{L}_{O}(P^-_u, N_u)$
        }
        \Else{
            $\mathcal{L} = \mathcal{L}_{F}(P^+_u, N_u)$ + $\mathcal{L}_{F}(P^-_u, N_u)$
        }
        Update student model $f$
    }
}
\caption{Training Procedure of \proposed}
\label{algo:tp}
\end{algorithm}

\noindent
\subsection{The Overall Training Process}
\label{subsec:tp}
Algorithm \ref{algo:tp} summarizes the whole training process.
The knowledge construction is applied in a personalized manner for each user $u$.
Also, the knowledge construction and $\mathcal{L}_{O}$ are applied until the student learns from the final converged teachers (i.e., $v[x]$ equals $E$, $\forall x$).
We conduct the knowledge construction every $p$ epoch, since changing the target permutation every epoch is time-consuming and unnecessary.
In this work, we set $p$ as 10.
A detailed analysis of \proposed's offline training costs is provided in Appendix \ref{app:cost}.

\section{Experiments}

\label{sec:experimentsetup}

\noindent
\textbf{Experiment settings.}
We use three real-world datasets: Amazon-music (Amusic), CiteULike, and Foursquare.
We randomly divide each user’s interactions into train/valid/test sets in a 60\%/20\%/20\% split \cite{CML}.
We use two top-$K$ ranking metrics: Recall@$K$ (R@$K$) and NDCG@$K$ (N@$K$)\footnote{Here, NDCG is computed for implicit feedback. 
To avoid confusion, we use NDCG to refer to the recommendation accuracy in the experiment section.}.
For the student model, we use MF, ML, and DNN \cite{DERRD, DCD, BD}. 
We set the user/item embedding dimension (or the bottleneck size for autoencoder) as 64 for all teacher models and 6 for~all~student~models so that the student has roughly one-tenth of the learning parameters used by the teacher as done in \cite{DERRD, DCD, BD}.

\vspace{0.1cm}
\noindent
\textbf{Baseline methods.}
We compare \proposed with various distillation methods.
All compared methods except MTD transfer the ensemble ranking of the converged heterogeneous teachers (denoted as \textbf{Ensemble}) which is generated in the same way as \proposed.
The first group of baselines includes a point-wise KD approach.
\vspace{-\topsep}
\begin{itemize}[leftmargin=*]
    \item \textbf{RD \cite{RD}} transfers the importance of top-ranked items. The importance is defined by each item's ranking position.
\end{itemize}
\vspace{-\topsep}
\noindent
The second group includes ranking matching KD methods (Sec.\ref{sec:rankingKD}).
\begin{itemize}[leftmargin=*]
\vspace{-\topsep}
    \item \textbf{RRD \cite{DERRD}} is a ranking matching method proposed for the recommender system. It uses listwise loss focusing on top-ranked~items.
    
    \item \textbf{MTD (Multi-Teacher Distillation)}: 
    We note that multi-teacher KD methods in other fields \cite{MT_KD1, ONE, MT_KD6, zhu2020ensembled} commonly use the trainable importance of each teacher on each data instance.
    Borrowing the idea, we develop MTD that transfers the knowledge with the trainable importance of each teacher on each user's ranking.
\end{itemize}
\vspace{-\topsep}
\noindent
The third group includes the state-of-the-art KD methods for ranking that use advanced schemes to improve the distillation~quality.
\begin{itemize}[leftmargin=*] \vspace{-\topsep}
    \item \textbf{CL-DRD \cite{CL-DRD}} is the state-of-the-art KD method for document retrieval. 
    It applies curriculum learning where the learning difficulty is predefined by the absolute ranking position.

    \item \textbf{DCD \cite{DCD}} is the state-of-the-art KD method for recommender system.
    It uses the dual correction loss, which corrects what the student has failed to accurately predict, along with RRD.
\end{itemize}

\vspace{-\topsep}
\noindent
Note that KD methods \cite{KD, BD, zhu2020ensembled, xia2022device} that directly use predicted scores are not applicable, as the output score distributions of the teachers and the student are very different in our task.

\vspace{-0.15cm}

\subsection{Distillation Effects Comparison}
\label{sec:result}
Table \ref{tab:main_b} presents the recommendation performance of the student models trained by different KD methods.
Also, Table \ref{tab:D_b} summarizes the discrepancy values of the best baseline (i.e., DCD) and \proposed.
`Best Teacher' denotes the teacher model showing the best performance among all heterogeneous teachers on each dataset.
\vspace{-\topsep}
\begin{itemize}[leftmargin=*]
\item For all datasets and student models, \proposed significantly outperforms all baselines, effectively improving the performance of the student\footnote{We also tried KD from various teachers (e.g., Best Teacher/Homogeneous teacher ensemble). \proposed improves the distillation efficacy in all cases (Appendix \ref{app:sup}).}.
Further, in terms of the number of recommended items ($K$), \proposed shows larger improvements for R@10/N@10 compared to R@50/N@50, which is advantageous for real-world services that aim to provide the most preferred items to their users.
Also, Table \ref{tab:D_b} shows that \proposed indeed achieves considerably lower discrepancy compared to the~best~baseline.

\item Compared to the point-wise KD approach (i.e., RD), the other KD methods that directly transfer the ranking orders consistently show higher performances, which again shows the importance of ranking knowledge in the top-$K$ recommendation.
On the one hand, MTD shows limited performance compared to RRD. 
Due to the high sparsity of interaction data, the user-wise learnable importance can be easily overfitted.

\item Ranking KD methods with advanced schemes (i.e., CL-DRD, DCD) improve the distillation effectiveness to some extent. 
However, for CL-DRD, we observe considerable variations for each dataset and student model.
One possible reason is that it uses predefined rules for defining the difficulty and controlling the difficulty level without considering the student's state.
On the other hand, DCD directly transfers what the student model has failed to predict, achieving consistently higher performance than~RRD.

\item Table \ref{tab:inference_time} presents the number of parameters and inference latency of the ensemble and \proposed.
We increase the size of the student (MF) until it achieves comparable performance to the ensemble.
Compared to the ensemble that incurs high inference costs due to the multiple model forwards, \proposed can significantly reduce the costs by distilling knowledge into the compact student model.
\end{itemize}
\vspace{-\topsep}

\noindent
\textbf{Further comparison on model generalization.}
As the easy-to-hard learning paradigm is known to yield a more generalizable model \cite{jiang2015self}, we further assess how accurately the student model captures the user preferences when the knowledge from the teachers is severely limited.
We randomly split the set of users into two sets with 80\%/20\% ratio (i.e., $\mathcal{U}_{g_1}, \mathcal{U}_{g_2}$).
Then, we train all teacher models by using only 80\% of training interactions of $\mathcal{U}_{g_1}$,
i.e., the teachers have limited knowledge of $\mathcal{U}_{g_1}$ and are not aware of $\mathcal{U}_{g_2}$.
Finally, we train the student model by using all training data with the distillation on $\mathcal{U}_{g_1}$\footnote{For training interactions not used for the teachers' training, we use the original~CF~loss.}.
That is, the student model is required to learn from the data by itself with incomplete guidance from the teachers.
In Table \ref{tab:main_g}, we observe that \proposed achieves significant improvements compared to DCD in all cases.
These results show that the student model trained by \proposed has a better generalization ability and can find more accurate hidden preferences by itself.~
We believe this can be advantageous in settings where new interactions are constantly being added, and we leave the study of applying \proposed to the continual learning \cite{Continual} in the future~work.

\begin{table*}[ht]
\caption{The recommendation performance comparison. 
$Imp$ denotes the improvement of \proposed over the best baseline.}
\label{tab:main_b}
\footnotesize
\renewcommand{\arraystretch}{0.7}
\renewcommand{\tabcolsep}{1.2mm}
\centering
\resizebox{0.99\linewidth}{!}{
\begin{tabular}{c|c|cccc|cccc|cccc}
\hline
& \multirow{2}{*}{\textbf{Method}} & \multicolumn{4}{c|}{\textbf{Amusic}} & \multicolumn{4}{c|}{\textbf{CiteULike}} & \multicolumn{4}{c}{\textbf{Foursquare}} \\
\cline{3-14}
  &  & \textbf{R@10} & \textbf{N@10} & \textbf{R@50} & \textbf{N@50} & \textbf{R@10} & \textbf{N@10} & \textbf{R@50} & \textbf{N@50} & \textbf{R@10} & \textbf{N@10} & \textbf{R@50} & \textbf{N@50} \\
\hline\hline
  & Best Teacher & 0.0972 & 0.0706 & 0.2475 & 0.1139  & 0.1337 & 0.0994 & 0.2844 & 0.1392 & 0.1147 & 0.1085 & 0.2723 & 0.1635\\
  & Ensemble& 0.1096 & 0.0820 & 0.2719 & 0.1273 & 0.1550 & 0.1156 & 0.3144 & 0.1571 & 0.1265 & 0.1213 & 0.2910 & 0.1786  \\
  \hline
\multirow{8}{*}{\begin{tabular}[c]{@{}c@{}}Student   \\      Model:\\      MF\end{tabular}} & w/o KD & 0.0449 & 0.0303 & 0.1451 & 0.0594 & 0.0568 & 0.0422 & 0.1372 & 0.0634 & 0.0726 & 0.0666 & 0.1806 & 0.1047 \\
  & RD & 0.0522 & 0.0387 & 0.1602 & 0.0693 & 0.0610 & 0.0472 & 0.1514 & 0.0725 & 0.0778 & 0.0703 & 0.1921 & 0.1153  \\
  & RRD & 0.0890 & 0.0659 & 0.2353 & 0.1077  & 0.0973 & 0.0740 & 0.2422 & 0.1113 & 0.0982 & 0.0905 & 0.2539 & 0.1446\\
  & MTD & 0.0901 & 0.0649 & 0.2279 & 0.1043  & 0.0993 & 0.0749 & 0.2425 & 0.1118 & 0.0955 & 0.0890 & 0.2402 & 0.1394\\
  & CL-DRD & 0.0883 & 0.0648 & 0.2375 & 0.1071  & 0.1033 & 0.0794 & 0.2512 & 0.1175 & 0.1001 & 0.0933 & 0.2528 & 0.1464\\
  & DCD & 0.0956 & 0.0675 & 0.2380 & 0.1079  & 0.1106 & 0.0851 & 0.2640 & 0.1246 & 0.1034 & 0.0965 & 0.2547 & 0.1491\\
  & \proposed & \textbf{0.1036}* & \textbf{0.0747}* & \textbf{0.2469}* & \textbf{0.1157}*  & \textbf{0.1379}* & \textbf{0.1031}* & \textbf{0.2814}* & \textbf{0.1396}* & \textbf{0.1118}* & \textbf{0.1036}* & \textbf{0.2722}* & \textbf{0.1594}*\\
\cline{2-14}
  & \textit{Imp}  & 8.37\% & 10.67\% & 3.74\% & 7.23\%  & 24.68\% & 21.15\% & 6.59\% & 12.04\% & 8.12\% & 7.36\% & 6.87\% & 6.91\%\\
\hline
\multirow{8}{*}{\begin{tabular}[c]{@{}c@{}}Student   \\      Model:\\      ML\end{tabular}} & w/o KD  & 0.0447 & 0.0310 & 0.1522 & 0.0623 & 0.0210 & 0.0148 & 0.0859 & 0.0323 & 0.0184 & 0.0139 & 0.0804 & 0.0356 \\
  & RD & 0.0706 & 0.0507 & 0.1874 & 0.0840 & 0.0835 & 0.0615 & 0.1914 & 0.0890 & 0.0729 & 0.0677 & 0.1811 & 0.1059\\
  & RRD & 0.0903 & 0.0643 & 0.2422 & 0.1074  & 0.0981 & 0.0701 & 0.2529 & 0.1116 & 0.0925 & 0.0813 & 0.2505 & 0.1366\\
  & MTD & 0.0843 & 0.0590 & 0.2293 & 0.1003  & 0.0944 & 0.0690 & 0.2519 & 0.1098 & 0.0909 & 0.0811 & 0.2440 & 0.1347\\
  & CL-DRD & 0.0866 & 0.0621 & 0.2409 & 0.1061  & 0.0989 & 0.0718 & 0.2583 & 0.1131 & 0.0931 & 0.0825 & 0.2541 & 0.1387\\
  & DCD & 0.0928 & 0.0653 & 0.2466 & 0.1086 & 0.1003 & 0.0724 & 0.2592 & 0.1144 & 0.0943 & 0.0845 & 0.2530 & 0.1399 \\
  & \proposed & \textbf{0.1020}* & \textbf{0.0751}* & \textbf{0.2470} & \textbf{0.1156}* & \textbf{0.1251}* & \textbf{0.0916}* & \textbf{0.2686}* & \textbf{0.1287}* & \textbf{0.1039}* & \textbf{0.0962}* & \textbf{0.2645}* & \textbf{0.1521}* \\
\cline{2-14}
  & \textit{Imp} & 9.91\% & 15.01\% & 0.16\% & 6.45\% & 24.73\% & 26.52\% & 3.63\% & 12.50\% & 10.18\% & 13.85\% & 4.09\% & 8.72\% \\
\hline
\multirow{8}{*}{\begin{tabular}[c]{@{}c@{}}Student   \\      Model:\\      DNN\end{tabular}} & w/o KD & 0.0460 & 0.0324 & 0.1396 & 0.0597 & 0.0414 & 0.0339 & 0.1095 & 0.0518 & 0.0693 & 0.0665 & 0.1608 & 0.0987 \\
  & RD & 0.0531 & 0.0378 & 0.1545 & 0.0670  & 0.0584 & 0.0445 & 0.1440 & 0.0671 & 0.0746 & 0.0683 & 0.1820 & 0.1060\\
  & RRD & 0.0851 & 0.0613 & 0.2255 & 0.1016  & 0.1034 & 0.0792 & 0.2552 & 0.1186 & 0.1016 & 0.0939 & 0.2584 & 0.1484\\
  & MTD & 0.0802 & 0.0563 & 0.2210 & 0.0958 & 0.0982 & 0.0710 & 0.2322 & 0.1058 & 0.0888 & 0.0797 & 0.2321 & 0.1305  \\
  & CL-DRD & 0.0889 & 0.0623 & 0.2365 & 0.1047 & 0.1083 & 0.0816 & 0.2575 & 0.1183 & 0.1039 & 0.0983 & 0.2635 & 0.1536 \\
  & DCD & 0.0919 & 0.0646 & 0.2404 & 0.1071  & 0.1114 & 0.0838 & 0.2668 & 0.1240 & 0.1060 & 0.1017 & 0.2671 & 0.1576\\
  & \proposed & \textbf{0.1045}* & \textbf{0.0768}* & \textbf{0.2534}* & \textbf{0.1190}*  & \textbf{0.1381}* & \textbf{0.1050 }*& \textbf{0.2864}* & \textbf{0.1413}* & \textbf{0.1136}* & \textbf{0.1079 }*& \textbf{0.2759}* & \textbf{0.1642}*\\
\cline{2-14}
  & \textit{Imp}& 13.71\% & 18.89\% & 5.41\% & 11.11\% & 23.97\% & 25.30\% & 7.35\% & 13.95\% & 7.17\% & 6.10\% & 3.29\% & 4.19\% \\
\hline
\end{tabular}
}
\small{$*$ denotes significance from the paired t-test (0.05 level) against the best baseline.}
\vspace{-0.2cm}
\end{table*}

\begin{table*}[ht]
\caption{The recommendation performance comparison of DCD and \proposed on the generalization setup. 
}
\label{tab:main_g}
\footnotesize
\renewcommand{\arraystretch}{0.85}
\renewcommand{\tabcolsep}{1.2mm}
\centering
\resizebox{0.99\linewidth}{!}{
\begin{tabular}{c|c|c|cccc|cccc|cccc}
\hline
& \textbf{User} & \multirow{2}{*}{\textbf{Method}} & \multicolumn{4}{c|}{\textbf{Amusic}} & \multicolumn{4}{c|}{\textbf{CiteULike}} & \multicolumn{4}{c}{\textbf{Foursquare}}\\
\cline{4-15}
& \textbf{group} &  & \textbf{R@10} & \textbf{N@10} & \textbf{R@50} & \textbf{N@50} & \textbf{R@10} & \textbf{N@10} & \textbf{R@50} & \textbf{N@50} & \textbf{R@10} & \textbf{N@10} & \textbf{R@50} & \textbf{N@50} \\
\hline\hline
 & \multirow{2}{*}{$\mathcal{U}_{g_1}$} & Best Teacher & 0.0741 & 0.0562 & 0.2075 & 0.0948 & 0.1065 & 0.0789 & 0.2276 & 0.1111 & 0.1024 & 0.0977 & 0.2401 & 0.1457 \\
& & Ensemble & 0.0972 & 0.0691 & 0.2344 & 0.1089 & 0.1304 & 0.0955 & 0.2805 & 0.1346 & 0.1141 & 0.1079 & 0.2655 & 0.1601 \\
\hline
\multirow{8}{*}{\begin{tabular}[c]{@{}c@{}}Student \\      Model:\\      MF\end{tabular}} & \multirow{4}{*}{$\mathcal{U}_{g_1}$} & w/o KD & 0.0452 & 0.0307 & 0.1460 & 0.0600 & 0.0551 & 0.0430 & 0.1351 & 0.0642 & 0.0736 & 0.0683 & 0.1811 & 0.1063\\
 &  & DCD & 0.0784 & 0.0548 & 0.2126 & 0.0933 & 0.0955 & 0.0724 & 0.2256 & 0.1056 & 0.0984 & 0.0937 & 0.2317 & 0.1401 \\
 &  & \proposed & \textbf{0.0903}* & \textbf{0.0661}* & \textbf{0.2313}* & \textbf{0.1059}* & \textbf{0.1096}* & \textbf{0.0834}* & \textbf{0.2547}* & \textbf{0.1206}* & \textbf{0.1050 }*& \textbf{0.0993}* & \textbf{0.2483}* & \textbf{0.1494}* \\
 \cline{3-15}
 &  & \textit{Imp} & 15.18\% & 20.62\% & 8.80\% & 13.50\% & 14.76\% & 15.19\% & 12.90\% & 14.20\% & 6.71\% & 5.98\% & 7.16\% & 6.64\%  \\
 \cline{2-15}
 & \multirow{4}{*}{$\mathcal{U}_{g_2}$} & w/o KD & 0.0434 & 0.0286 & 0.1411 & 0.0570 & 0.0555 & 0.0427 & 0.1389 & 0.0649 & 0.0728 & 0.0676 & 0.1808 & 0.1058 \\
 &  & DCD & 0.0567 & 0.0370 & 0.1706 & 0.0699  & 0.0722 & 0.0517 & 0.1785 & 0.0805 & 0.0738 & 0.0717 & 0.1839 & 0.1087\\
 &  & \proposed & \textbf{0.0668}* & \textbf{0.0465}* & \textbf{0.1841}* & \textbf{0.0788}*  & \textbf{0.0801}* & \textbf{0.0558}* & \textbf{0.2008}* & \textbf{0.0876}* & \textbf{0.0880}* & \textbf{0.0829}* & \textbf{0.2088}* & \textbf{0.1227}*\\
 \cline{3-15}
 &  & \textit{Imp} & 17.81\% & 25.68\% & 7.91\% & 12.73\% & 10.94\% & 7.93\% & 12.49\% & 8.82\% & 19.24\% & 15.62\% & 13.54\% & 12.88\% \\
 \hline
\multirow{8}{*}{\begin{tabular}[c]{@{}c@{}}Student \\      Model:\\      ML\end{tabular}} & \multirow{4}{*}{$\mathcal{U}_{g_1}$} & w/o KD & 0.0442 & 0.0308 & 0.1518 & 0.0623 & 0.0209 & 0.0148 & 0.0856 & 0.0322 & 0.0201 & 0.0158 & 0.0860 & 0.0386 \\
 &  & DCD & 0.0817 & 0.0585 & 0.2179 & 0.0970 & 0.0861 & 0.0618 & 0.2214 & 0.0974 & 0.0891 & 0.0799 & 0.2289 & 0.1288\\
 &  & \proposed & \textbf{0.0908}* & \textbf{0.0641}* & \textbf{0.2284}* & \textbf{0.1027}*  & \textbf{0.1110}* & \textbf{0.0831}* & \textbf{0.2445}* & \textbf{0.1173}* & \textbf{0.1025}* & \textbf{0.0960}* & \textbf{0.2499}* & \textbf{0.1475}*\\
 \cline{3-15}
 &  & \textit{Imp} & 11.14\% & 9.57\% & 4.82\% & 5.88\%  & 28.92\% & 34.47\% & 10.43\% & 20.43\% & 15.04\% & 20.15\% & 9.17\% & 14.52\%\\
 \cline{2-15}
 & \multirow{4}{*}{$\mathcal{U}_{g_2}$} & w/o KD & 0.0468 & 0.0318 & 0.1537 & 0.0627 & 0.0208 & 0.0147 & 0.0869 & 0.0324 & 0.0201 & 0.0156 & 0.0861 & 0.0384 \\
 &  & DCD & 0.0532 & 0.0368 & 0.1601 & 0.0680 & 0.0522 & 0.0371 & 0.1644 & 0.0665 & 0.0631 & 0.0562 & 0.1826 & 0.0976 \\
 &  & \proposed & \textbf{0.0651}* & \textbf{0.0453}* & \textbf{0.1906}* & \textbf{0.0805}* & \textbf{0.0806}* & \textbf{0.0560}* & \textbf{0.2012}* & \textbf{0.0868}* & \textbf{0.0709}* & \textbf{0.0626}* & \textbf{0.2052}* & \textbf{0.1099}* \\
 \cline{3-15}
 &  & \textit{Imp} & 22.37\% & 23.10\% & 19.05\% & 18.38\%  & 54.41\% & 50.94\% & 22.38\% & 30.53\% & 12.36\% & 11.39\% & 12.38\% & 12.60\%\\
 \hline
\multirow{8}{*}{\begin{tabular}[c]{@{}c@{}}Student \\      Model:\\      DNN\end{tabular}} & \multirow{4}{*}{$\mathcal{U}_{g_1}$} & w/o KD & 0.0463 & 0.0329 & 0.1398 & 0.0602  & 0.0408 & 0.0335 & 0.1088 & 0.0514 & 0.0691 & 0.0665 & 0.1606 & 0.0986\\
 &  & DCD & 0.0805 & 0.0579 & 0.2118 & 0.0947 & 0.0882 & 0.0675 & 0.2177 & 0.1010 & 0.0914 & 0.0854 & 0.2231 & 0.1314 \\
 &  & \proposed & \textbf{0.0902}* & \textbf{0.0664}* & \textbf{0.2308}* & \textbf{0.1063}* & \textbf{0.1106}* & \textbf{0.0841}* & \textbf{0.2527}* & \textbf{0.1206}* & \textbf{0.1014}* & \textbf{0.0960}* & \textbf{0.2472}* & \textbf{0.1438}* \\
 \cline{3-15}
 &  & \textit{Imp} & 12.05\% & 14.68\% & 8.97\% & 12.25\% & 25.40\% & 24.59\% & 16.08\% & 19.41\% & 10.94\% & 12.41\% & 10.80\% & 9.44\%  \\
 \cline{2-15}
 & \multirow{4}{*}{$\mathcal{U}_{g_2}$} & w/o KD & 0.0450 & 0.03034 & 0.1385 & 0.05795  & 0.0435 & 0.0352 & 0.1121 & 0.0532 & 0.0698 & 0.0671 & 0.1615 & 0.0993\\
 &  & DCD & 0.0501 & 0.0342 & 0.1551 & 0.0642  & 0.0586 & 0.0440 & 0.1516 & 0.0673 & 0.0670 & 0.0621 & 0.1686 & 0.0974\\
 &  & \proposed & \textbf{0.0647}* & \textbf{0.0448}* & \textbf{0.1829}* & \textbf{0.0751}*  & \textbf{0.0780}* & \textbf{0.0572}* & \textbf{0.1901}* & \textbf{0.0833}* & \textbf{0.0820}* & \textbf{0.0778}* & \textbf{0.2001}* & \textbf{0.1188}*\\
 \cline{3-15}
 &  & \textit{Imp}& 29.14\% & 30.99\% & 17.92\% & 16.98\% & 33.11\% & 30.00\% & 25.40\% & 23.77\% & 22.39\% & 25.28\% & 18.68\% & 21.97\% \\
 \hline
\end{tabular}
}
\small{$*$ denotes significance from the paired t-test (0.05 level) against the best baseline.}
\vspace{-0.2cm}
\end{table*}

\begin{table}[ht!]
\centering
\caption{Discrepancy comparison of DCD and \proposed.}
\label{tab:D_b}
\renewcommand{\arraystretch}{0.8}
\resizebox{1.01\linewidth}{!}{
\begin{tabular}{cc|cc|cc|cc}
\hline
\multirow{2}{*}{} & \multirow{2}{*}{\textbf{Method}} & \multicolumn{2}{c|}{\textbf{Amusic}} & \multicolumn{2}{c|}{\textbf{CiteULike}} & \multicolumn{2}{c}{\textbf{Foursquare}} \\ \cline{3-8}
 &  & \textbf{D@10} &\textbf{ D@50} & \textbf{D@10} & \textbf{D@50} & \textbf{D@10} & \textbf{D@50} \\
 \hline\hline
\multirow{4}{*}{\rotatebox{90}{MF}} & w/o KD & 0.8984 & 0.8028 & 0.7668 & 0.6608 & 0.6211 & 0.4625 \\
 & DCD & 0.6610 & 0.5286 & 0.4433 & 0.3184 & 0.2856 & 0.1690 \\
 & \proposed & 0.5929 & 0.4709 & 0.3401 & 0.2359 & 0.2135 & 0.1210 \\ \cline{2-8}
 & \textit{Imp}& 10.30\% & 10.92\% & 23.28\% & 25.91\% & 25.25\% & 28.40\% \\ \hline
\multirow{4}{*}{\rotatebox{90}{ML}} & w/o KD & 0.9072 & 0.8149 & 0.9583 & 0.8936 & 0.9532 & 0.8654 \\
 & DCD & 0.6699 & 0.5283 & 0.5272 & 0.3659 & 0.3429 & 0.1967 \\
 & \proposed & 0.5967 & 0.4756 & 0.3856 & 0.2693 & 0.2653 & 0.1481 \\ \cline{2-8}
 & \textit{Imp} & 10.93\% & 9.98\% & 26.86\% & 26.40\% & 22.63\% & 24.71\% \\ \hline
\multirow{4}{*}{\rotatebox{90}{DNN$\,\,\,$}} & w/o KD & 0.9070 & 0.8232 & 0.8225 & 0.7330 & 0.6438 & 0.5119 \\
 & DCD & 0.6862 & 0.5462 & 0.4836 & 0.3427 & 0.2377 & 0.1315 \\
 & \proposed & 0.5338 & 0.4170 & 0.2942 & 0.2101 & 0.2014 & 0.1110\\ \cline{2-8}
 & \textit{Imp} & 22.21\% & 23.65\% & 39.16\% & 38.69\% & 15.27\% & 15.59\% \\\hline
\end{tabular}}
\end{table}

\begin{table}[ht!]
\caption{Accuracy-efficiency trade-off. Time (s) indicates the average wall time for generating each user's recommendation. We use PyTorch with CUDA from RTX A5000 and Xeon Gold 6226R CPU.}
\label{tab:inference_time}
\small
\centering
\renewcommand{\tabcolsep}{0.85mm}
\renewcommand{\arraystretch}{0.9}
\resizebox{1.0\linewidth}{!}{
\begin{tabular}{c|c|cc|cc}
\hline
\multirow{2}{*}{\textbf{Dataset}} & \multirow{2}{*}{\textbf{Method}} & \multicolumn{2}{c|}{\textbf{Accuracy}}  & \multicolumn{2}{c}{\textbf{Efficiency}} \\\cline{3-6}
 & & \textbf{R@10} & \textbf{N@10} & \textbf{\#Params (emb.size)} & \textbf{Time} \\
 \hline\hline
\multirow{2}{*}{Amusic} & Ensemble & 0.1096 & 0.0820 & 5.79M$\,\,\,$ (64) & 10.57s \\ 
 & \proposed & 0.1102 & 0.0817 & 0.27M$\,\,\,$ (18) & 0.82s \\
 \hline
\multirow{2}{*}{CiteULike} & Ensemble & 0.1550 & 0.1156 & 11.72M (64) & 22.10s \\
 & \proposed & 0.1548 & 0.1150 & 0.45M$\,\,\,$ (15) & 1.10s \\
 \hline
\multirow{2}{*}{Foursquare} & Ensemble & 0.1265 & 0.1213 & 18.52M (64) & 35.47s \\
 & \proposed & 0.1263 & 0.1214 & 0.96M$\,\,\,$ (20) & 2.12s\\
 \hline
\end{tabular}}
\end{table}

\subsection{Study of \proposed}
\label{subsec:studyH}
We provide in-depth analysis to provide a deeper insight of \proposed.
Supplementary results including the hyperparameter study and the ablation study are provided in Appendix \ref{app:sup}.

\subsubsection{\textbf{Why are teachers' training trajectories helpful?}}
To get clues to the question, we analyze what knowledge is revealed from the teachers' intermediate states (i.e., E1-E4, E4: the converged state).
All reported results correspond to the average value from all teachers for each intermediate state.
For intuitive understanding, we report the relative ratio to the value from E4.

\begin{figure}[t]
\vspace{-0.12cm}
\centering
\hspace{-0.15cm}
\includegraphics[width=0.47\linewidth]{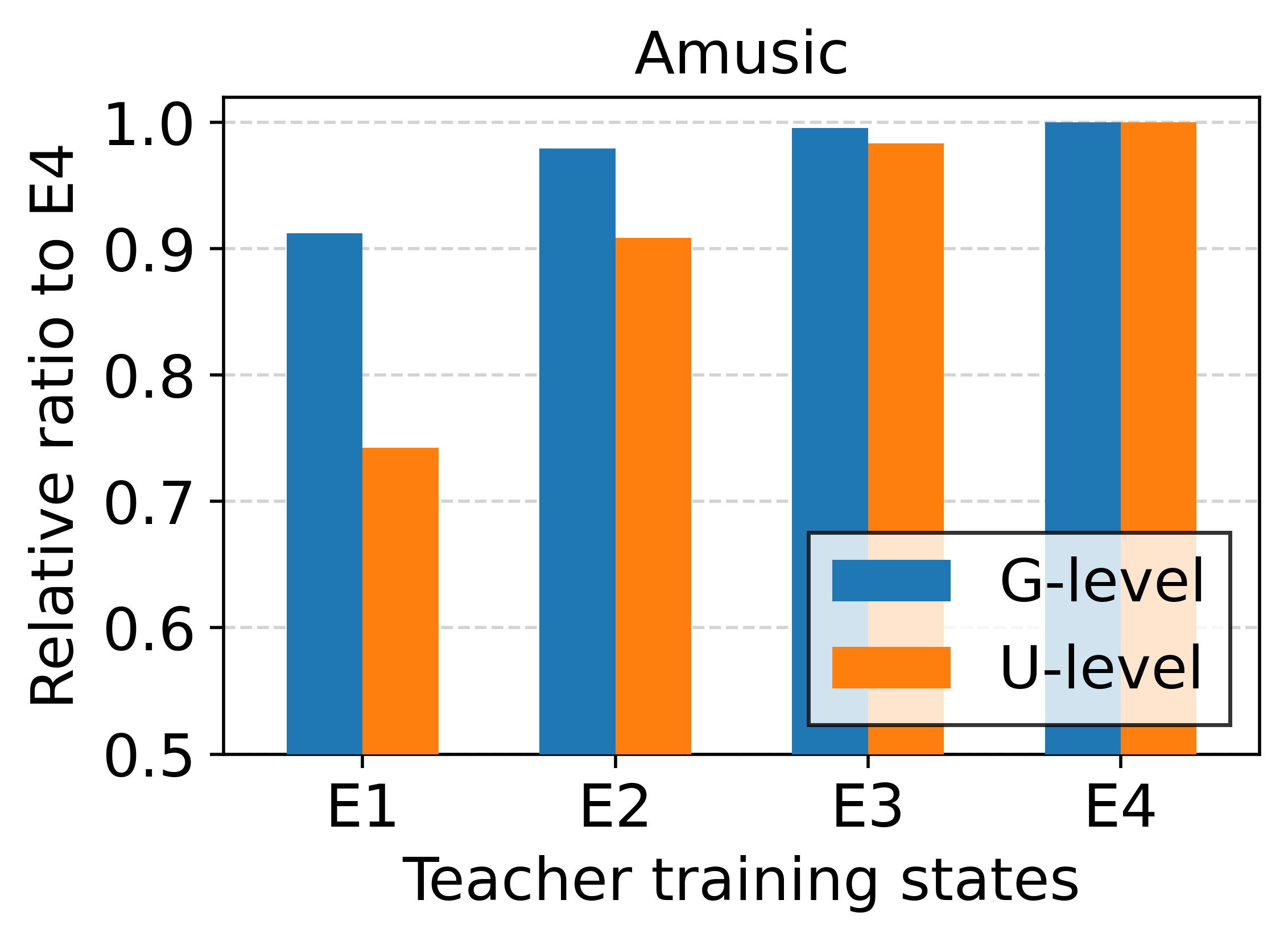}
    \hspace{-0.2cm}
    \includegraphics[width=0.47\linewidth]{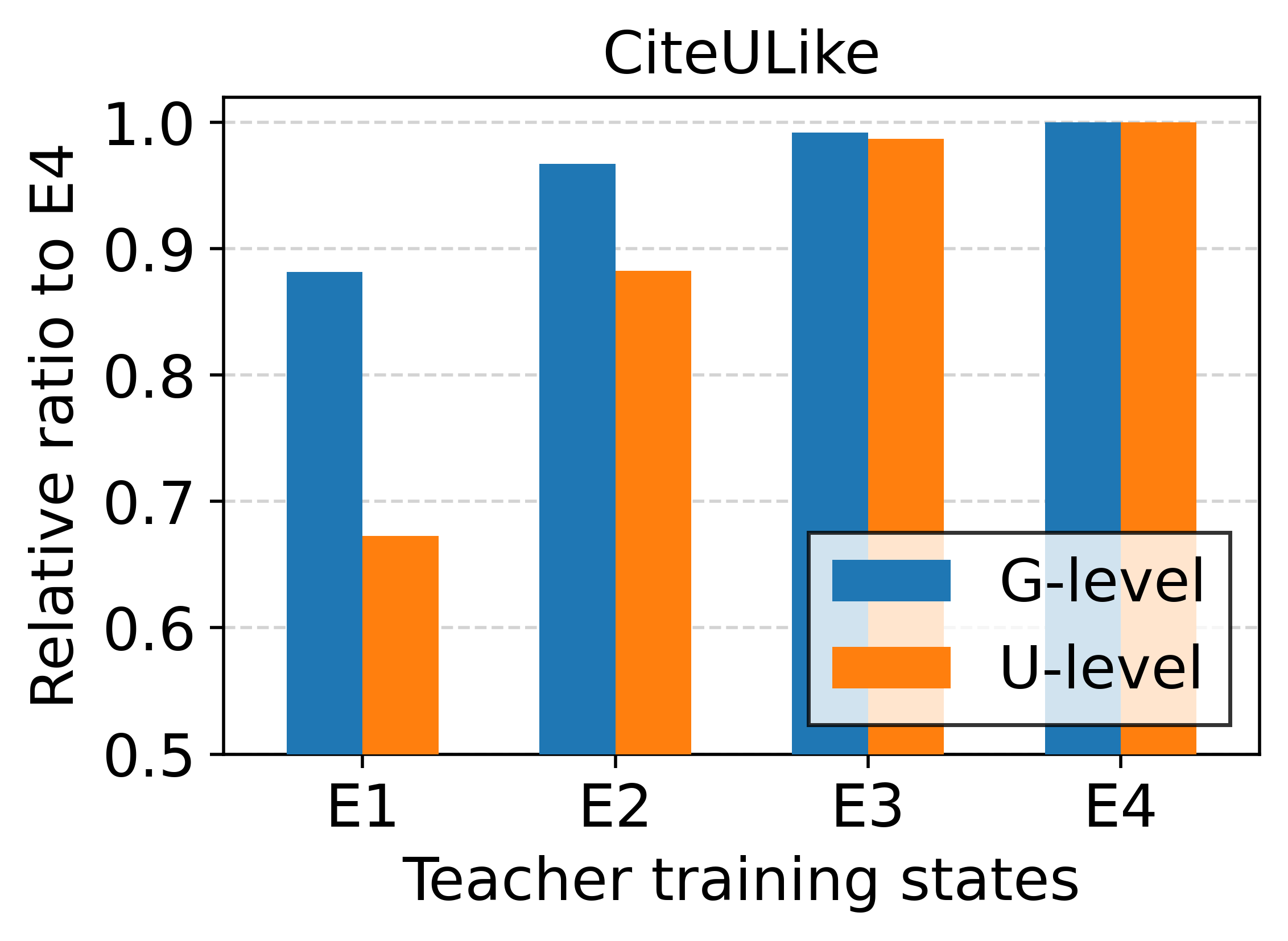}\hspace{-0.1cm}
    \caption{The relative ratio of the capability to capture group-level and user-level preference (Metric: R@50).}
    \label{fig:PG_level}
    \Description{The relative ratio of the capability to capture group-level and user-level preference (Metric: R@50).}
    \vspace{-0.1cm}
\end{figure}

\begin{itemize}[leftmargin=*]
    \item We investigate the items included in the ranking list from each intermediate state.
    In the union of all users' top-50 ranking lists, we count (1) how many unique items exist and (2) how many items belong to unpopular items.\footnote{We regard items with the lowest 30\% of interaction numbers as unpopular items.}
    Table \ref{tab:unique_unpop} presents the ratios of the number of unique items and unpopular items from each teacher training state.
    We observe that the items included in the recommendations are gradually diversified during the training.
    In addition, the proportions of unpopular items, which reflect more personalized preferences than popular items, are also progressively increasing.

    \item We analyze each state's capability of capturing the group-level and user-level preferences.
    To assess the group-level preference, we identify 50 user groups by $k$-means on interaction history, then find top-50 test items frequently consumed by each group.
    We use the group-level test items for the evaluation of each user in the group.
    The user-level preference corresponds to the original recommendation task.
    We use R@50 for evaluating the recommendation accuracy of both levels.
    Figure \ref{fig:PG_level} summarizes the ratios of the group-level and user-level preferences from each teacher training state.
    We observe that the early states capture a limited user-level preference, but a fairly accurate group-level preference.
    That is, although teachers' earlier predictions include less personalized (fine-grained) preferences, they reveal overall group-level (coarse-grained) patterns.
\end{itemize}

\noindent
To sum up, in \proposed, by using the teachers' intermediate states, the items making up the knowledge are gradually changed so that it progressively reveals more diverse and personalized preferences.
This aligns well with the idea of curriculum learning that first focuses on the overall concept and gradually learns specific patterns \cite{curriculum}.
It is worth noting that all our baselines transfer \textit{fixed knowledge} throughout the student's training.

\subsubsection{\textbf{Study of dynamic knowledge construction (DKC)}}
\begin{figure}[t]
\centering
\hspace{-0.2cm}
    \includegraphics[width=0.51\linewidth]{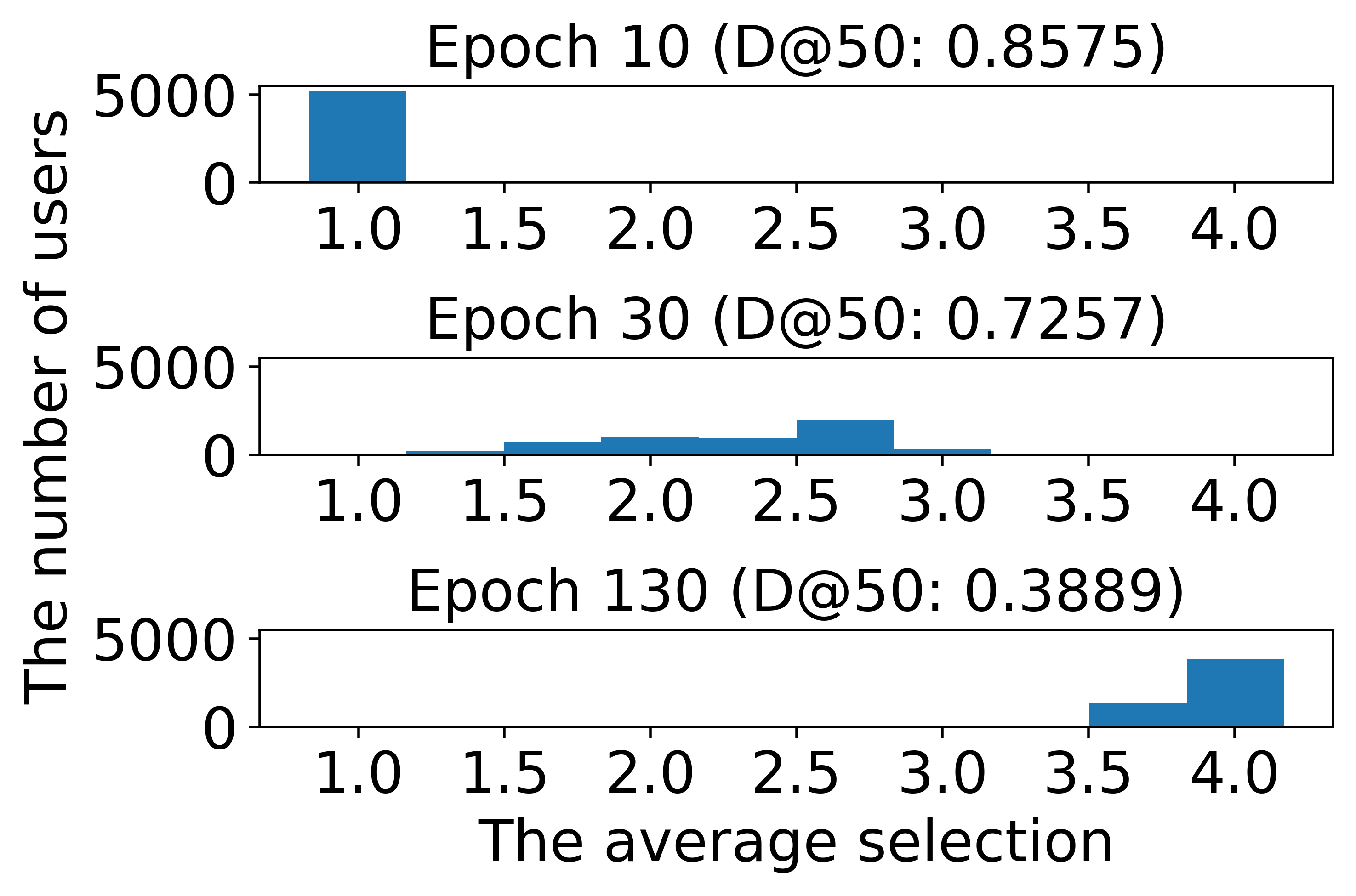}
    \hspace{-0.2cm}
    \includegraphics[width=0.49\linewidth]{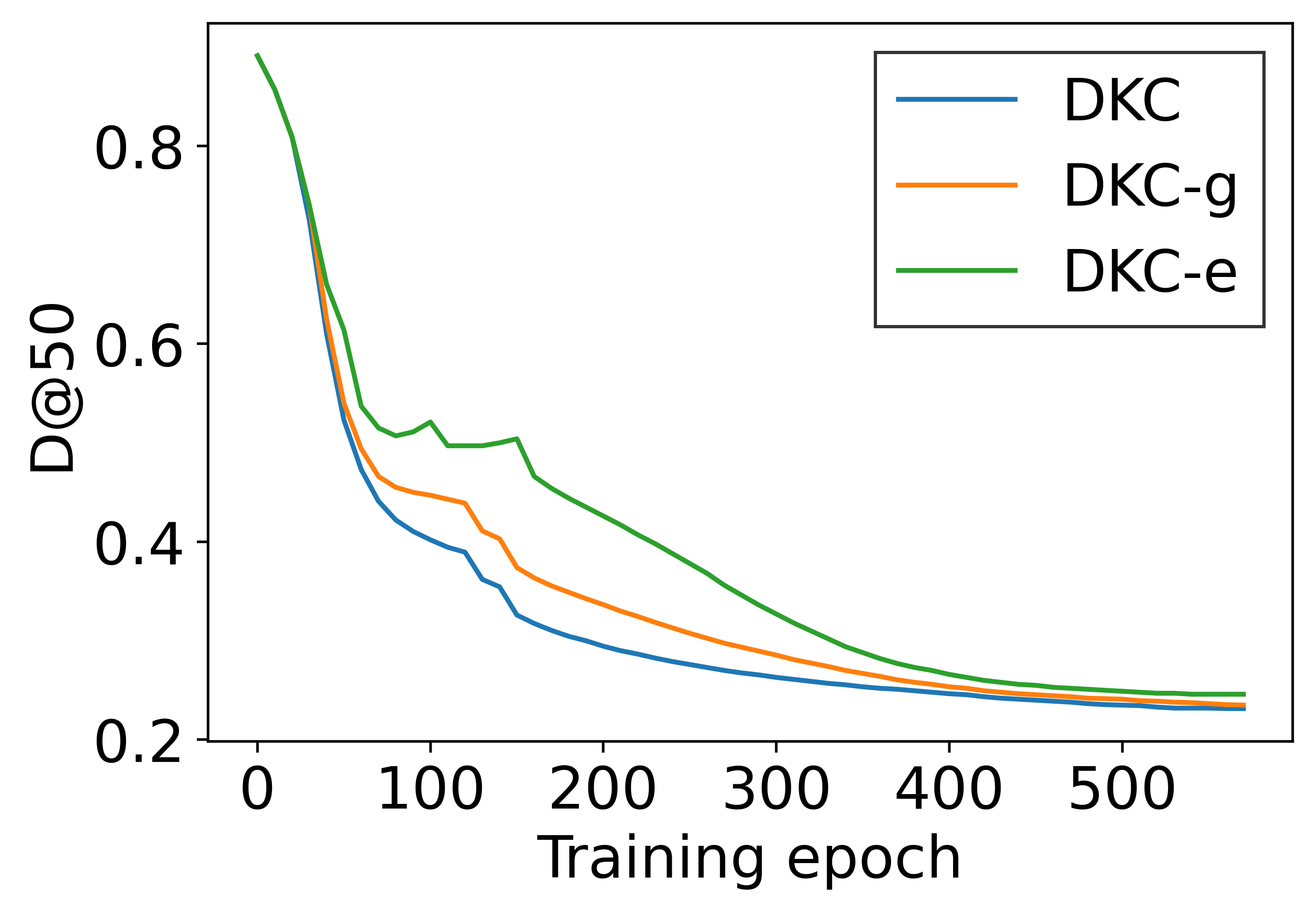}\hspace{-0.1cm}
    \caption{
    (left) the average selection distributions at epoch 10/30/130 of the student model.
    (right) $D@50$ curves with variants of dynamic knowledge construction. The discrepancy is computed to the ensemble ranking from converged teacher models.}
    \label{fig:DKC_anal}
\Description{(left) the average selection distributions at epoch 10/30/130 of the student. (right) $D@50$ curves with variants of dynamic knowledge construction. The discrepancy is computed to the ensemble ranking from converged teachers}
\end{figure}

We provide a detailed analysis of DKC.
Here, we use MF (student) on CiteULike dataset.
Fig.\ref{fig:DKC_anal}(left) presents how users' average knowledge selections (i.e., $\mathbb{E}_x[v_u[x]]$) change.
During the student's training, the discrepancy to the final ranking is lowered by learning earlier predictions, and \proposed gradually moves toward more difficult knowledge following the teachers' trajectories.
Also, the knowledge selection differs for each user, which reflects different learning difficulties for each user's knowledge.
In Fig.\ref{fig:DKC_anal}(right), we compare two variants of DKC:
`DKC-g' uses a global selection variable computed by averaging user-wise selection,
`DKC-e' uses a simple rule that moves to the next state after a certain epoch interval.
The interval is determined by equally dividing the epoch when DKC ends by 3.
Compared to the variants, DKC provides more efficient curricula for the student model.
DKC-e rather increases the discrepancy (around epoch 100) because it cannot reflect the student's learning state.
These results show the benefits of considering both varying user-wise learning difficulties and the learning status of the student model in DKC.


\begin{table}[t]
\centering
\caption{The relative ratios of the number of unique items and unpopular items in top-50 ranking lists from the teacher models.}
\label{tab:unique_unpop}
\renewcommand{\arraystretch}{0.7}
\resizebox{0.99\linewidth}{!}{
\begin{tabular}{c|cc|cc}
\hline
\textbf{Teacher} & \multicolumn{2}{c|}{\textbf{Amusic}} & \multicolumn{2}{c}{\textbf{CiteULike}} \\ \cline{2-5}
\textbf{training states} & \multicolumn{1}{l}{\textbf{unique}} & \multicolumn{1}{l|}{\textbf{unpopular}} & \multicolumn{1}{l}{\textbf{unique}} & \multicolumn{1}{l}{\textbf{unpopular}} \\ \hline \hline
E1  & 0.8386 & 0.7205 & 0.7055 & 0.7190 \\
E2  & 0.8863 & 0.8618 & 0.8207 & 0.8366 \\
E3  & 0.9479 & 0.9254 & 0.9230 & 0.9191 \\
E4  & 1.0 & 1.0 & 1.0 & 1.0 \\
\hline
\end{tabular}}
\end{table}

\section{Conclusion}
\label{sec:conclusion}
We propose a new \proposed framework to compress the ensemble knowledge of heterogeneous recommendation models into a lightweight model, so as to reduce the huge inference costs while retaining high accuracy.
From our analysis, we show that distillation from heterogeneous teachers is particularly challenging and teachers' training trajectories can help to ease such high learning difficulties.
Based on the idea of easy-to-hard learning, \proposed uses \textit{dynamic knowledge construction} to provide progressively difficult ranking knowledge and \textit{adaptive knowledge transfer} to gradually transfer finer-grained ranking information.
We provide extensive experiments showing that \proposed significantly improves the distillation efficacy and the generalization of the student model.
Based on its great compatibility with existing models, we expect that our \proposed framework can be a solution for the accuracy-efficiency trade-off of the recommender system.

\pagebreak
\newpage
\clearpage

\noindent \textbf{Acknowledgments.}
We thank Sang-Wook Kim, Minsu Cho, Dongwoo Kim, and Chanyoung Park for their insightful feedback.
This work was supported by Microsoft Research Asia and IITP (No.2022-00155958), IITP grant funded by MSIT (No.2018-0-00584, 2019-0-01906), NRF grant funded by MSIT (No.2020R1A2B5B03097210).
\noindent \textbf{DOI.} \url{https://doi.org/10.5281/zenodo.7594882}

\bibliographystyle{ACM-Reference-Format}         
\bibliography{acmart}

\pagebreak
\newpage
\clearpage
\label{sec:appendix}
\appendix
\clearpage

\section{Appendix}

\subsection{Experiment Setup}
The source code of HetComp is publicly available through the author’s GitHub repository\footnote{\url{https://github.com/SeongKu-Kang/HetComp_WWW23}}.

\vspace{-0.1cm}
\label{sub:setup}
\subsubsection{\textbf{Dataset}}
We use three real-world datasets: Amazon-music (Amusic), CiteULike, and Foursquare.
These datasets are publicly accessible and also widely used in previous work \cite{BD, DERRD, DCD, BUIR, SSCDR}.
We follow the preprocessing of \cite{BUIR} (CiteULike, Foursquare) and apply 10-core filtering (Amusic).
Table \ref{tbl:datastats} provides the data statistics.

\vspace{-0.2cm}
\begin{table}[h]
\centering
\renewcommand{\arraystretch}{0.7}
\caption{Statistics of the datasets.}
\begin{tabular}{c|c|c|c|c}
\hline
\textbf{Dataset} & \textbf{User \#} & \textbf{Item \#} & \textbf{Interaction \#} & \textbf{Density} \\\hline \hline
Amazon-music & 5,729 & 9,267 & 65,344 & 0.001231\\
CiteULike & 5,219 & 25,181 & 125,580 & 0.000956 \\
Foursquare & 19,465 & 28,593 & 1,115,108 & 0.002004 \\
\hline
\end{tabular}
\label{tbl:datastats}
\vspace{-0.5cm}
\end{table}

\subsubsection{\textbf{Experiment details}}
For all experiments and inferences, we use PyTorch with CUDA from RTX A5000 and Xeon Gold 6226R CPU.
We report the average value of five independent runs.
For all baselines, we use the public implementations provided by the authors.
However, as done in \cite{CL-DRD}, we found that their sampling processes for top-ranked unobserved items (i.e., $P^-$) are unnecessary, and removing the processes gave considerable performance improvements for the ranking matching KD methods.
For this reason, we remove the sampling process for all ranking matching methods in our experiments.
In D$@K$, $\lambda$ that controls the sharpness of the exponential function is set to 10.
For dynamic knowledge construction, we use D$@50$, $p$ is set to 10, and $\alpha$ is set to $1.05$. 
As the student model gradually converges, we adopt a simple annealing schedule that decreases the value $\alpha=\alpha \times 0.995$ every $p$ epoch.
We set $E$ as 4, each of which corresponds to the checkpoint at 25\%, 50\%, 75\%, and 100\% of the converged epoch for each teacher.
Lastly, $|P^-|$ is set to 50.
For baseline-specific hyperparameters, we tune them in the ranges suggested by~the~original~papers.

\subsection{Study on the Ensemble of Teacher Models}
\label{sec:app_MTS}

\subsubsection{\textbf{Recommendation models for teacher}}
\label{sec:ranking_teacher}
In this work, we use six recommendation models with different architectures and learning objectives.
These models are the representative models for each model type and have shown competitive performance.
\begin{itemize}[leftmargin=*] \vspace{-\topsep}
    \item \textbf{MF} (BPR \cite{BPR}): a matrix factorization-based model trained by a pair-wise ranking loss. The ranking score is defined by the inner product of the user and item latent factors.
    \item \textbf{ML} (CML \cite{CML}): a metric learning-based model trained by a pair-wise hinge loss. The ranking score is defined by the Euclidean distance in the unit-ball metric space.
    \item \textbf{DNN} (NeuMF \cite{NeuMF}): a deep neural network-based model trained by binary cross-entropy. The ranking score is computed by the non-linear function of multi-layer perceptrons.
    \item \textbf{GNN} (LightGCN \cite{he2020lightgcn}): a graph neural network-based model~trained by a pair-wise loss. 
    The ranking score is computed by aggregating the  user and item representations from multiple~GNN~layers.
    \item \textbf{AE} (VAE \cite{VAE}): a variational autoencoder-based model. The ranking score is computed by the generative module (i.e., decoder).
    \item \textbf{I-AE}: a variant of VAE that learns item-side interactions \cite{autorec, jointAE}. It is known that the item-side autoencoder captures complementary aspects to its user-side counterpart \cite{jointAE}.
\end{itemize} \vspace{-\topsep}
\noindent
The best ensemble performance is achieved by using all six models, and we provide empirical evidence supporting our configuration in the next subsection.

\vspace{-0.1cm}
\subsubsection{\textbf{Necessity of heterogeneous teachers}}
Table \ref{tab:ensemble_study} presents an empirical study on the configuration of the teacher models.
`Best Teacher' denotes the teacher model showing the best performance on each dataset.
`Ensemble' denotes the ensemble results of six heterogeneous models.
`Ensemble-id' denotes the ensemble of six identical models having different initialization\footnote{We use the model showing the best performance on each dataset.}. 
`w/o model' denotes the ensemble of five models excluding the model~from~`Ensemble'.

First, Ensemble consistently shows higher performance than Ensemble-id and the best teacher model.
We investigate the correlations of model predictions for Ensemble and Ensemble-id\footnote{We compare test interactions included in the top-50 ranking of each model.
We compute the correlation of all pairs of models for each user and compare the average value.
We use the Matthews correlation coefficient (MCC) provided in sklearn.}.
The former shows 23\% (Amusic), 20\% (CiteULike), and 38\% (Foursquare) lower prediction correlations compared to the latter.
It is well known that models with high diversity boost the overall ensemble accuracy, and the lower correlations of Ensemble support its higher performance to some extent.
This observation also aligns with the recent studies of the ensemble \cite{rank_aggregation} and the multi-teacher KD \cite{MT_KD2, MT_KD4, MT_KD6} showing that the diversity of models is a key factor of performance.
Lastly, the best performance of Ensemble is achieved by consolidating all six models; all cases of five models (i.e., w/o model) show limited performance compared to Ensemble.


\begin{table*}[thp]
\caption{Ensemble Study. The best performance is achieved by using six heterogeneous models.}
\label{tab:ensemble_study}
\footnotesize
\renewcommand{\arraystretch}{0.65}
\centering
\resizebox{0.99\linewidth}{!}{
\begin{tabular}{c|cccc|cccc|cccc}
\hline
\multirow{2}{*}{Method} & \multicolumn{4}{c|}{\textbf{Amusic}} & \multicolumn{4}{c|}{\textbf{CiteULike}} & \multicolumn{4}{c}{\textbf{Foursquare}} \\
\cline{2-13}
 & \textbf{R@10} & \textbf{N@10} & \textbf{R@50} & \textbf{N@50} & \textbf{R@10} & \textbf{N@10} & \textbf{R@50} & \textbf{N@50} & \textbf{R@10} & \textbf{N@10} & \textbf{R@50} & \textbf{N@50} \\
 \hline\hline
Best Teacher & 0.0972 & 0.0706 & 0.2475 & 0.1139 & 0.1337 & 0.0994 & 0.2844 & 0.1392 & 0.1147 & 0.1085 & 0.2723 & 0.1635 \\
Ensemble & 0.1096 & 0.0820 & 0.2719 & 0.1273 & 0.1550 & 0.1156 & 0.3144 & 0.1571 & 0.1265 & 0.1213 & 0.2910 & 0.1730 \\
\hline
Ensemble-id & 0.1013 & 0.0736 & 0.2569 & 0.1181 & 0.1511 & 0.1130 & 0.2952 & 0.1505 & 0.1215 & 0.1174 & 0.2853 & 0.1709 \\ \hline
w/o MF & 0.1091 & 0.0826 & 0.2634 & 0.1266 & 0.1545 & 0.1161 & 0.3128 & 0.1564 & 0.1206 & 0.1124 & 0.2883 & 0.1708 \\
w/o ML & 0.1050 & 0.0798 & 0.2574 & 0.1232 & 0.1541 & 0.1137 & 0.3085 & 0.1541 & 0.1222 & 0.1181 & 0.2854 & 0.1746 \\
w/o DNN & 0.1074 & 0.0819 & 0.2642 & 0.1263 & 0.1543 & 0.1147 & 0.3134 & 0.1560 & 0.1193 & 0.1099 & 0.2855 & 0.1678 \\
w/o GNN & 0.1085 & 0.0823 & 0.2634 & 0.1259 & 0.1544 & 0.1150 & 0.3117 & 0.1557 & 0.1197 & 0.1124 & 0.2857 & 0.1703 \\
w/o AE & 0.1032 & 0.0776 & 0.2520 & 0.1204 & 0.1482 & 0.1101 & 0.3046 & 0.1508 & 0.1177 & 0.1102 & 0.2775 & 0.1658 \\
w/o I-AE & 0.1040 & 0.0797 & 0.2613 & 0.1243 & 0.1519 & 0.1129 & 0.3046 & 0.1528 & 0.1204 & 0.1150 & 0.2812 & 0.1710 \\ \hline
\end{tabular}
}
\vspace{-0.4cm}
\end{table*}


\vspace{-0.1cm}
\subsubsection{\textbf{Ensemble technique}}
\label{app:ensemble_technique}
We now describe our choice of the ensembling function, $g$ for ranking knowledge construction.
Since each teacher model better predicts certain user-item interactions than others, it is vital to selectively reflect their knowledge into the ensemble.
We note that \textit{the consistency of model prediction} is a key factor revealing the reliability of the prediction \cite{tc-ssl}. 
This factor has been successfully employed for RS for obtaining reliable negative samples \cite{NS_std} and for consolidating multiple heads of the multi-task learning model \cite{concf}.
More sophisticated techniques can be considered, but we empirically obtain satisfactory performance with our choice.
We provide comparisons with the technique using trainable importance in the experiments.

$g$ generates an ensemble ranking $\pi^\text{d}$ by consolidating a set of permutations $\Pi=[\pi^1, \pi^2, ..., \pi^M]$.
For top-$K$ of each permutation $\pi^x$, each ranking prediction $r(\pi^x,i)$ has an importance score $c^x_{i}$:
\begin{equation}
\begin{aligned}
c^x_{i} = \exp(-r(\pi^x,i)/\lambda) + \exp(-\operatorname{std}[r(\pi^x,i)]/\lambda)
\end{aligned}
\end{equation}
The first term put the higher score on items with a higher ranking, and the second term uses the variance of predictions to favor items that the model makes consistent predictions.
Following \cite{NS_std}, we define the consistency based on the variance of the latest~5~epochs:
\begin{equation}
\begin{aligned}
\operatorname{std}\left[r(\pi^x,i)\right] =&\sqrt{\sum_{s=t-4}^{t}\left(\left[r(\pi^x,i)\right]_{s}-\operatorname{Mean}\left[r(\pi^x,i)\right]\right)^{2} / 5}, \\
\operatorname{Mean}\left[r(\pi^x,i)\right] =&\sum_{s=t-4}^{t}\left[r(\pi^x,i)\right]_{s} / 5 .
\end{aligned}
\end{equation}
where $\left[r(\pi^x,i)\right]_{s}$ denotes the ranking prediction at epoch $s$.
Finally, the ensemble ranking is generated by reranking items based on the overall importance, i.e., $\mathbb{E}_{x}[c^x_{i}]$.
Note that the consistency is precomputed only once before the distillation and incurs no additional costs for \proposed.

\subsection{Offline Training Cost of \proposed}
\label{app:cost}
\proposed requires additional space and computation costs mostly for the knowledge construction process.
For the teachers' training trajectories $\mathcal{T}$, we store the permutations of top-ranked ($K$) unobserved items.
Note that the rankings of the remaining unobserved items are unnecessary.
Also, we use the permutations of observed items.
In sum, \proposed uses $(K \times E)$ + $|P^+_u|$ space for user $u$ on each teacher.
$K$ and $|P^+_u|$ usually have small values in many recommendation scenarios,
and we empirically obtain satisfactory results with $E$ around 3 as long as they are well distributed (Figure~\ref{fig:alpha}).

\begin{table}[ht]
\vspace{-0.3cm}
\centering
\renewcommand{\arraystretch}{0.75}
\caption{The average time cost of each epoch during the offline training.}
\label{tab:epochtime}
\begin{tabular}{c|c|c|c}
\hline
\textbf{Method} & \textbf{Amusic} & \textbf{CiteULike} & \textbf{Foursquare} \\
\hline \hline
RRD & 0.69s & 2.27s & 8.12s \\
DCD & 1.38s & 3.74s & 12.01s \\
\proposed & 1.15s & 3.08s & 11.81s\\
\hline
\end{tabular}
\vspace{-0.3cm}
\end{table}

Table \ref{tab:epochtime} presents the average time cost of \proposed and ranking matching KD baselines.
We use MF as the student model, and similar results are also observed with other base models.
Compared to RRD which uses basic listwise learning, DCD and \proposed incur additional training costs as they require further computations for the distillation.
However, unlike DCD which requires such computation throughout the training, our knowledge construction mostly occurs at the earlier training of the student model.
As a result, \proposed shows lower average time costs compared to DCD.
Also, they show similar convergence behavior.
As shown in Section 5, \proposed can significantly improve the distillation quality by reducing the discrepancy to the teachers.
In this regard, \proposed can be considered an effective solution to reduce the online inference costs at the expense of some extra computations in offline training.

\subsection{Supplementary Experiment Results}
\label{app:sup}

\noindent
\textbf{Ablation study.}
Table \ref{tab:ablation} provides comparison with various ablations.
We report the results of MF (student) on the CiteULike dataset.
First, (a-b) shows that our two components (i.e., DKC and ADO) designed for the easy-to-hard learning effectively improve the student model.
Also, we compare diverse ways of transferring knowledge of observed items ($P^+$);
(c) shows that utilizing $P^+$ is indeed beneficial to improve the student, and (d), which corresponds to the naive approach transferring the whole item permutation, shows that ranking transfer without separating $P^+$ and $P^-$ has adverse effects as discussed in Sec.\ref{subsubsec:obs}.
Lastly, (e) shows that penalizing all unobserved items to have lower ranks than observed items is not effective.
These results support the superiority of our strategy that independently transfers $P^+$ and $P^-$.

\begin{table}[h]
\renewcommand{\arraystretch}{0.85}
\centering
 \caption{Ablation study. DKC: dynamic knowledge construction (Sec.\ref{subsec:dkc}), ADO: adaptive distillation objective (Sec.\ref{subsubsec:ado}).}
\label{tab:ablation}
\resizebox{1.01\linewidth}{!}{
\renewcommand{\tabcolsep}{0.75mm}
\begin{tabular}{l|cccc|c}
\hline
\textbf{Ablation} & \textbf{R@10} & \textbf{N@10} & \textbf{R@50} & \textbf{N@50} & \footnotesize{\textbf{\textit{Imp.}R@10}}\\
\hline \hline
\proposed & 0.1379 & 0.1031 & 0.2814 & 0.1396 & - \\
\hline
(a) w/o DKC (i.e., KD from Ensemble) & 0.1264 & 0.0951 & 0.2711 & 0.1308 & 9.10\%  \\
(b) w/o ADO (i.e., only $\mathcal{L}_F$) & 0.1311 & 0.0992 & 0.2765 & 0.1360 & 5.19\%\\
(c) w/o $\mathcal{L}_{KD}(P^+,N)$ & 0.1303 & 0.0994 & 0.2754 & 0.1349 & 5.83\%\\
(d) $\mathcal{L}_{KD}(\{P^+, P^-\}, N)$ & 0.1241 & 0.0903 & 0.2796 & 0.1314 & 11.12\%\\
(e) $\mathcal{L}_{KD}(P^+,\{P^-,N\}) + \mathcal{L}_{KD}(P^-,N)$ & 0.1262 & 0.0950 & 0.2665 & 0.1310 & 9.27\% \\
\hline
\end{tabular}}
\end{table}

\begin{figure}[h]
\centering
    \hspace{-0.2cm}
    \includegraphics[width=0.26\linewidth]{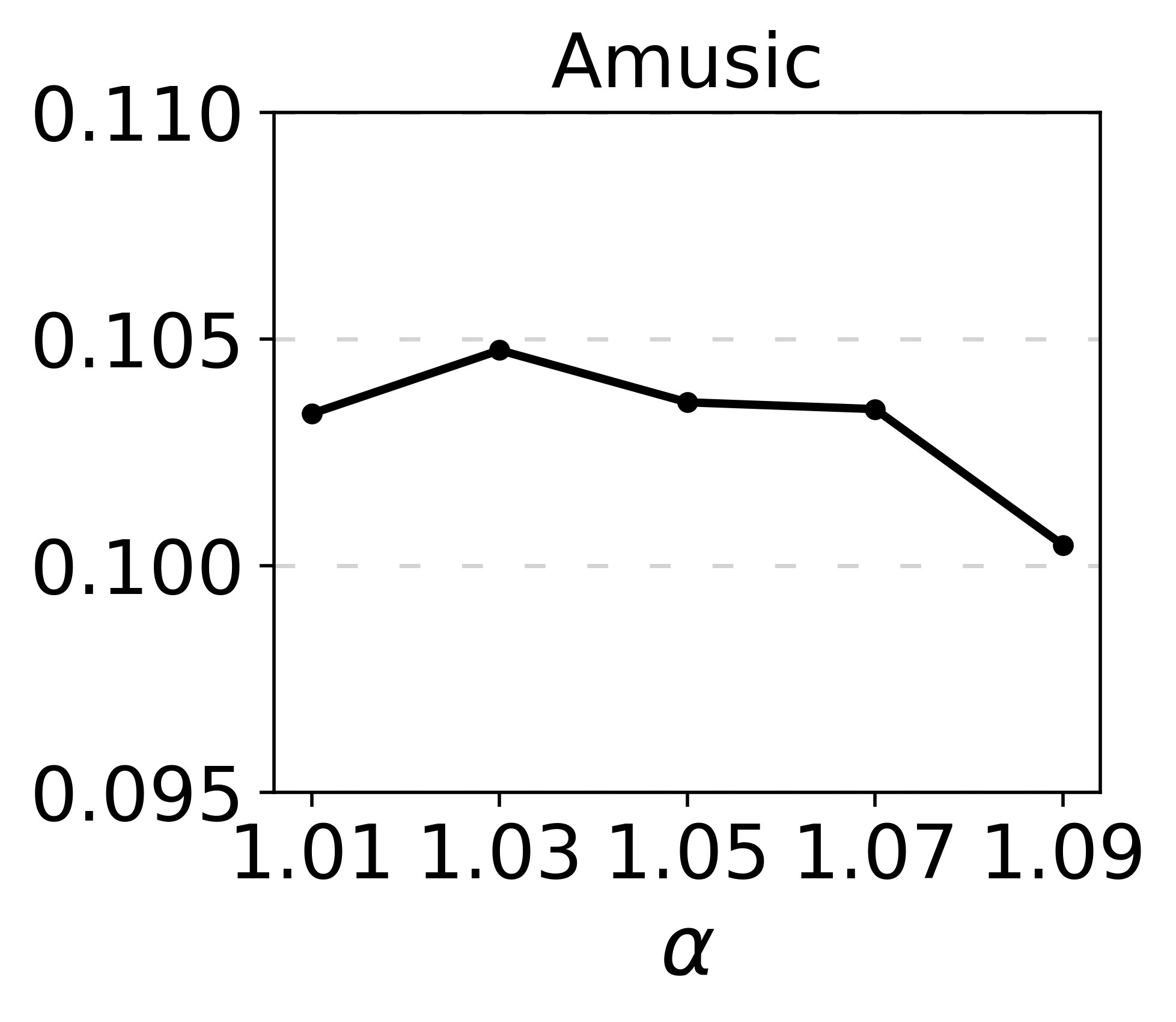}
    \hspace{-0.22cm}
    \includegraphics[width=0.26\linewidth]{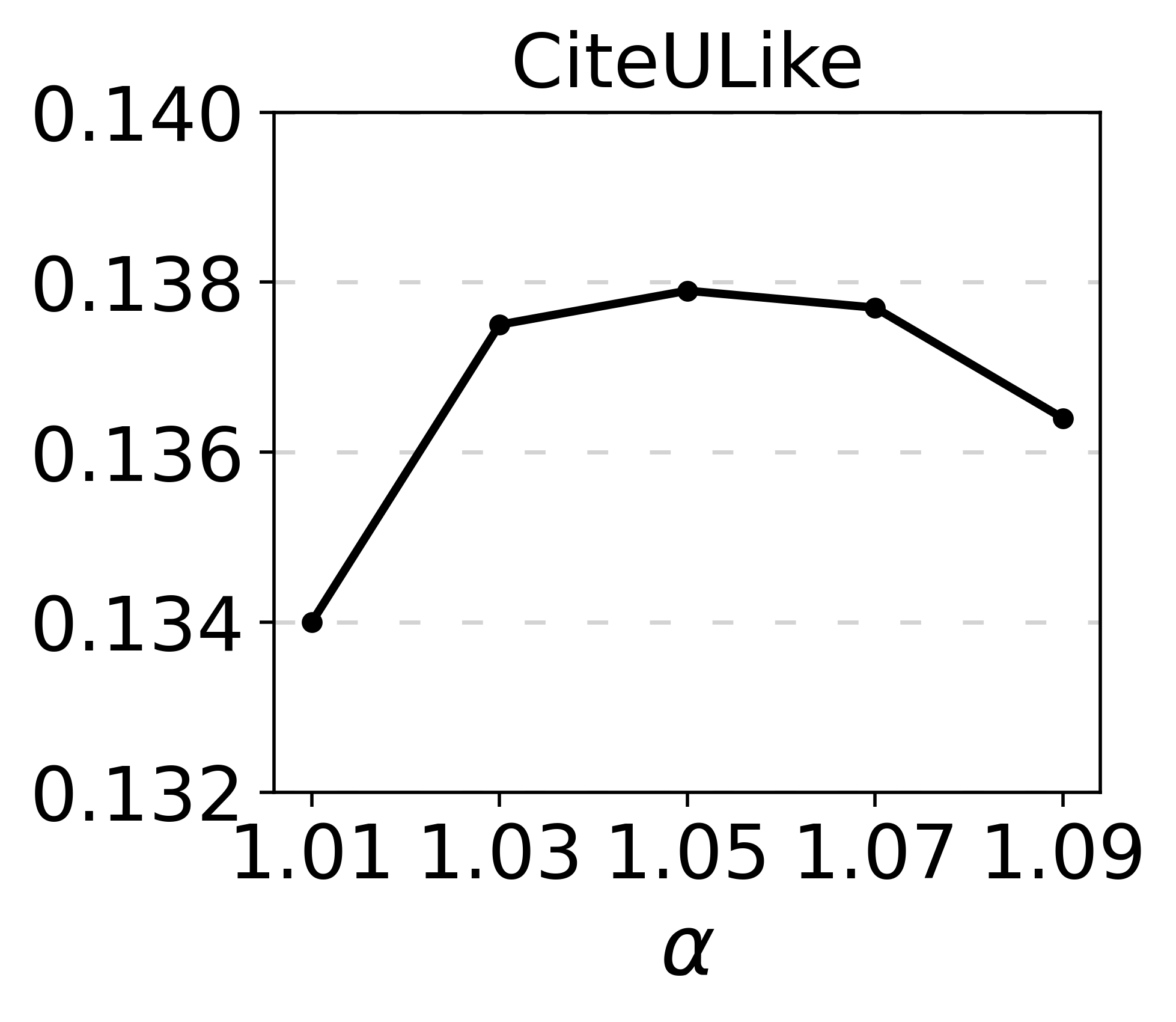}
    \hspace{-0.22cm}
    \includegraphics[width=0.26\linewidth]{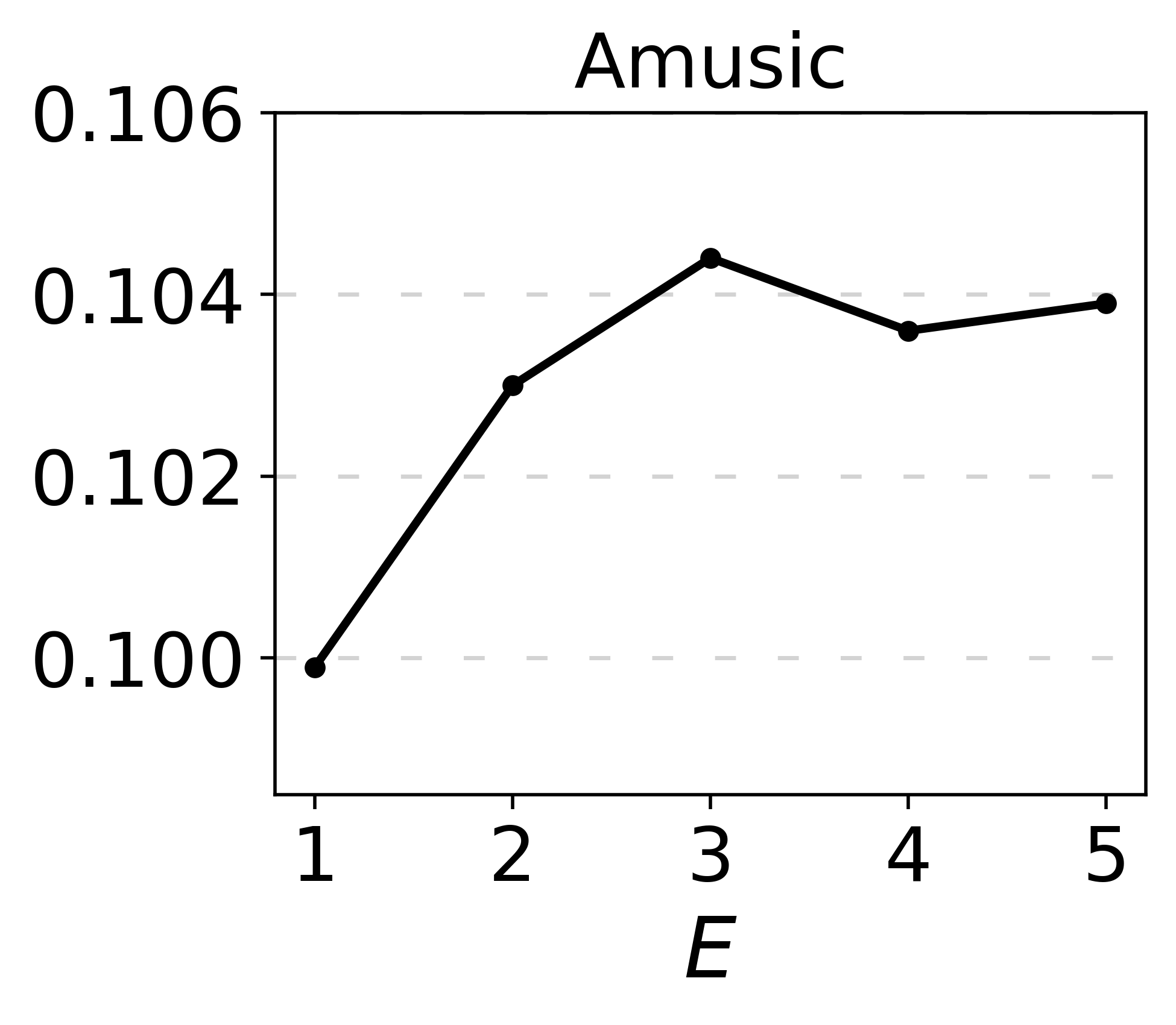}
    \hspace{-0.2cm}
    \includegraphics[width=0.26\linewidth]{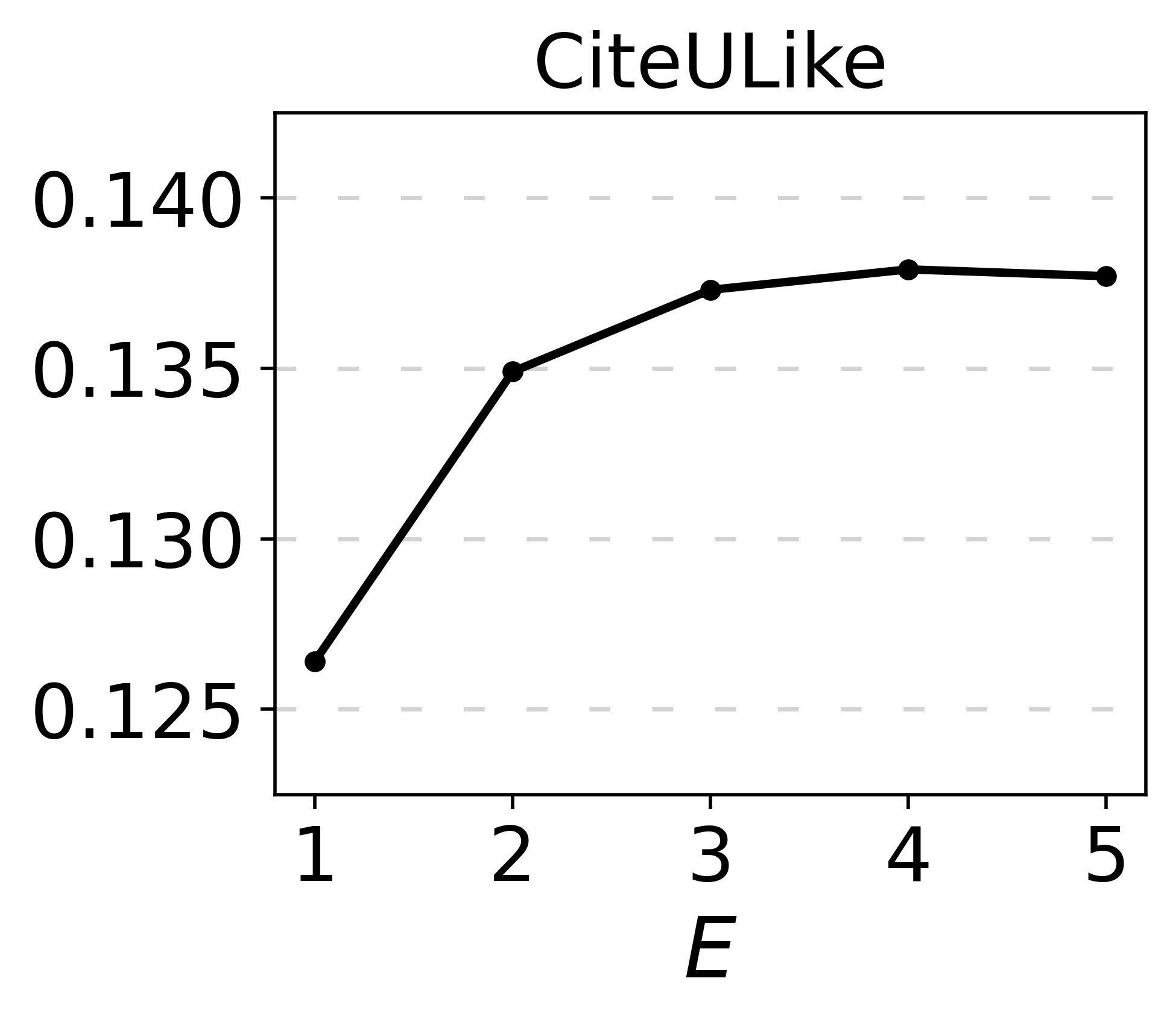}
    \hspace{-0.2cm}
    \caption{R@10 comparison with varying $\alpha$ and $E$.}
    \label{fig:alpha}
    \vspace{-0.3cm}
\end{figure}

\noindent
\textbf{Hyperparameter study.} Figure \ref{fig:alpha} presents the recommendation performance of \proposed (Student: MF) with varying $\alpha$ which controls the transition speed in the dynamic knowledge construction.
The best performance is observed at around $1.03$-$1.05$ on both datasets.

\noindent
\textbf{KD from various teachers.}
Table 11 presents the performance of DCD and \proposed when transferring knowledge of various~teachers.
We use MF (student) and the CiteULike dataset.
We observe that \proposed effectively improves the distillation quality and achieves the best recommendation performance in all settings, from (a-b) homogeneous model distillation to (c-d) cross-model distillation, and (e) distillation from the ensemble of heterogeneous models. 

\begin{table}[h]
\vspace{-0.2cm}
\caption{Performance comparison with various teachers.}
\centering
\label{tab:various}
\renewcommand{\tabcolsep}{1.2mm}
\renewcommand{\arraystretch}{0.8}
\resizebox{1\linewidth}{!}{
\begin{tabular}{c|cccc|c}
\hline
  & \textbf{R@10} & \textbf{N@10} & \textbf{R@50} & \textbf{N@50} & \textbf{D@10} \\
\hline \hline
(a)  Teacher: MF & 0.1249 & 0.0915 & 0.2604 & 0.1273 & - \\
 DCD & 0.0937 & 0.0698 & 0.2206 & 0.1044 & 0.3831 \\
 \proposed & 0.1092 & 0.0794 & 0.2373 & 0.1128 & 0.3247 \\ \hline
(b)  Teacher: Ensemble (MF) & 0.1395 & 0.1037 & 0.2763 & 0.1395 & - \\
 DCD & 0.1004 & 0.0749 & 0.2307 & 0.1088 & 0.4044 \\
 \proposed & 0.1194 & 0.0878 & 0.2584 & 0.1237 & 0.3286 \\ \hline
(c) Teacher: LightGCN & 0.1337 & 0.0994 & 0.2844 & 0.1392 & - \\
  DCD & 0.1041 & 0.0778 & 0.2367 & 0.1120 & 0.4164 \\
  \proposed & 0.1228 & 0.0896 & 0.2510 & 0.1230 & 0.3590 \\ \hline
(d) Teacher: Ensemble (LightGCN) & 0.1511 & 0.1130 & 0.2952 & 0.1505 & - \\
  DCD & 0.1108 & 0.0838 & 0.2495 & 0.1186 & 0.4180 \\
  \proposed & 0.1305 & 0.0973 & 0.2630 & 0.1255 & 0.3377 \\ \hline
(e)  Teacher: Ensemble & 0.1550 & 0.1156 & 0.3144 & 0.1571 & - \\ 
  DCD & 0.1106 & 0.0851 & 0.2640 & 0.1246 & 0.4433 \\
  \proposed & 0.1379 & 0.1031 & 0.2814 & 0.1396 & 0.3401\\ \hline
\end{tabular}}
\vspace{-0.4cm}
\end{table}

\end{document}